\pdfoutput=1
\documentclass[lettersize, journal]{IEEEtran}
\usepackage[utf8]{inputenc} 
\usepackage[T1]{fontenc}   
\usepackage{hyperref}       
\usepackage{enumitem}
\usepackage{url}            
\usepackage{booktabs}       
\usepackage{amsfonts}       
\usepackage{nicefrac}       
\usepackage{microtype}      
\usepackage{xcolor}         
\usepackage{amsmath}
\usepackage{bbm}
\usepackage{multirow}
\usepackage{colortbl}
\usepackage{siunitx}
\usepackage{courier}
\usepackage{amsmath,amsfonts}
\usepackage{algorithmic}
\usepackage{array}
\usepackage{textcomp}
\usepackage{stfloats}
\usepackage{url}
\usepackage{amssymb}
\usepackage{verbatim}
\usepackage{graphicx}
\usepackage[noadjust]{cite}
\usepackage{capt-of}
\usepackage{bm}

\usepackage{tikz}
\usetikzlibrary{positioning, calc, arrows.meta}
\usetikzlibrary{shapes.geometric}
\usetikzlibrary{fadings}
\usetikzlibrary{shadings, fadings, calc}
\usepackage{tikz-3dplot}

\newcommand{\calA}{\mathcal{A}}
\newcommand{\A}{\mathcal{A}}
\newcommand{\calS}{\mathcal{S}}
\renewcommand{\S}{\mathcal{S}}

\newcommand{\vin}{v_{x\to x'}}
\newcommand{\vout}{\Omega}

\newcommand{\calL}{\mathcal{L}}

\newcommand{\Npat}{{{N_\mathrm{pat}}}}
\newcommand{\Ndir}{{{N_\mathrm{dir}}}}

\newcommand{\diag}{\operatorname{diag}}

\newcounter{relctr} 
\everydisplay\expandafter{\the\everydisplay\setcounter{relctr}{0}} 

\AtBeginDocument{} 

\usepackage[switch]{lineno}

\ifCLASSOPTIONcompsoc
    \usepackage[caption=false, font=footnotesize, labelfont=sf, textfont=sf]{subfig}
\else
\usepackage[caption=false, font=footnotesize]{subfig}
\fi
\definecolor{myLightOrange}{RGB}{230, 159, 0}
\definecolor{myLightBlue}{RGB}{86, 180, 233}
\definecolor{myGreen}{RGB}{0, 158, 115}
\definecolor{myYellow}{RGB}{240, 228, 66}
\definecolor{myDarkBlue}{RGB}{0, 114, 178}
\definecolor{myDarkOrange}{RGB}{213, 94, 0}
\definecolor{myPurple}{RGB}{204, 121, 167}

\definecolor{1stcol}{HTML}{56B4E9}
\definecolor{2ndcol}{HTML}{FFADC9}

\newcolumntype{R}{>{\raggedleft\arraybackslash}p{9.5mm}}
\newcolumntype{X}[1]{>{\raggedleft\arraybackslash}p{#1}}
\newcolumntype{C}[1]{>{\centering\arraybackslash}p{#1}}
\newcolumntype{L}[1]{>{\raggedright\arraybackslash}p{#1}}

\title{Differentiable Acoustic Radiance Transfer}
\author{
Sungho~Lee, Matteo~Scerbo, Seungu~Han, Min~Jun~Choi, Kyogu~Lee, \IEEEmembership{Senior Member, IEEE}, \\ and Enzo~De~Sena, \IEEEmembership{Senior Member, IEEE}
\thanks{
Manuscript submitted 15 October 2025; revised 4 March 2026; accepted 10 April 2026.
This work was supported in part by the UK's Engineering and Physical Sciences Research Council (EPSRC) under Grant EP/V002554/1 (Scalable Room Acoustics Modeling) and Grant EP/X032914/1 (Challenges in Immersive Audio Technology).
This work was also partly supported by the National Research Foundation of Korea (NRF) grant funded by the Korean government (MSIT) [No. RS-2025-24683892], [No. RS-2024-00461617], and the Institute of Information \& communications Technology Planning \& Evaluation (IITP) grant funded by the Korea government(MSIT) [NO.RS-2021-II211343]
The associate editor coordinating the review of this manuscript was Jie Zhang \emph{(Corresponding Author: Kyogu Lee)}.

Sungho Lee, Seungu Han, Min Jun Choi, and Kyogu Lee are with the Department of Intelligence and Information, Graduate School of Convergence Science and Technology, Seoul National University, Seoul, Republic of Korea (e-mail: sh-lee@snu.ac.kr; hansw0326@snu.ac.kr; choimj21@snu.ac.kr; kglee@snu.ac.kr).
Kyogu Lee is also with the Interdisciplinary Program in Artificial Intelligence (IPAI), Seoul, South Korea.

Matteo Scerbo was with the Institute of Sound Recording (IoSR), University of Surrey, GU2 7XH, Guildford, U.K. 
He is now with the Technical University of Denmark (DTU), Kongens Lyngby, Denmark (e-mail: mattes@dtu.dk).

Enzo De Sena is with the Institute of Sound Recording (IoSR), University of Surrey, GU2 7XH, Guildford, U.K. (e-mail:
e.desena@surrey.ac.uk).

Implementation and extra supplementary materials are available at \url{https://github.com/sh-lee97/dart}

}
}

\begin{document}
\maketitle

\begin{abstract}
Geometric acoustics is an efficient framework for room acoustics modeling, governed by the canonical time-dependent rendering equation.
Acoustic radiance transfer (ART) solves the equation by discretization, modeling time- and direction-dependent energy exchange between surface patches with flexible material properties.
We introduce DART, an efficient, differentiable implementation of ART that enables gradient-based optimization of material properties.
We evaluate DART on a simpler variant of acoustic field learning that aims to predict energy responses for novel source-receiver configurations.
Experimental results demonstrate that DART generalizes better under sparse measurement scenarios than existing signal processing and neural network baselines, while maintaining simplicity and full interpretability.
We open-source our implementation.

\end{abstract}

\newcommand{\incomingcolor}{myDarkOrange}
\newcommand{\incidentcolor}{myDarkBlue}
\newcommand{\outgoingcolor}{myGreen}

\pgfmathsetmacro{\viewElev}{65}
\pgfmathsetmacro{\viewAzim}{30}    
\pgfmathsetmacro{\viewscale}{1.68}    
\newif\ifconeborders \conebordersfalse   
\newif\ifshowaxes \showaxesfalse          
\pgfmathsetmacro{\chx}{0}        
\pgfmathsetmacro{\chy}{2.0}
\pgfmathsetmacro{\chz}{2.5}
\pgfmathsetmacro{\cix}{3}      
\pgfmathsetmacro{\ciy}{2.0}
\pgfmathsetmacro{\ciz}{0}
\pgfmathsetmacro{\patchHY}{0.7}   
\pgfmathsetmacro{\patchHZh}{0.7}   
\pgfmathsetmacro{\patchHXi}{0.7}  
\pgfmathsetmacro{\coneT}{1.4}      
\pgfmathsetmacro{\coneR}{0.35}     
\pgfmathsetmacro{\farM}{1.7}         
\pgfmathsetmacro{\farT}{\farM*\coneT}
\pgfmathsetmacro{\farR}{\farM*\coneR}
\pgfmathsetmacro{\isqrt}{1/sqrt(2)}  
\pgfmathsetmacro{\coneLayers}{50}      
\pgfmathsetmacro{\coneAlpha}{0.005}    
\pgfmathsetmacro{\dlen}{sqrt((\cix-\chx)*(\cix-\chx)+(\ciy-\chy)*(\ciy-\chy)+(\ciz-\chz)*(\ciz-\chz))}
\pgfmathsetmacro{\hjAx}{(\cix-\chx)/\dlen}
\pgfmathsetmacro{\hjAz}{(\ciz-\chz)/\dlen}
\pgfmathsetmacro{\hjPx}{-(\hjAz)}
\pgfmathsetmacro{\hjPz}{\hjAx}
\pgfmathsetmacro{\ikAx}{\hjAx}
\pgfmathsetmacro{\ikAz}{-(\hjAz)}
\pgfmathsetmacro{\ikPx}{-(\ikAz)}
\pgfmathsetmacro{\ikPz}{\ikAx}
\pgfmathsetmacro{\ilAx}{-(\hjAx)}
\pgfmathsetmacro{\ilAz}{-(\hjAz)}
\pgfmathsetmacro{\ilPx}{-(\ilAz)}
\pgfmathsetmacro{\ilPz}{\ilAx}

\newcommand{\computecone}[6]{%
  \pgfmathsetmacro{\cBx}{\coneT*#3 + \coneR*\isqrt*#5}
  \pgfmathsetmacro{\cBz}{\coneT*#4 + \coneR*\isqrt*#6}
  \coordinate (#1_b1) at ($(#2)+(\cBx, {\coneR*\isqrt}, \cBz)$);
  \coordinate (#1_b2) at ($(#2)+(\cBx, {-\coneR*\isqrt}, \cBz)$);
  \pgfmathsetmacro{\cBx}{\coneT*#3 - \coneR*\isqrt*#5}
  \pgfmathsetmacro{\cBz}{\coneT*#4 - \coneR*\isqrt*#6}
  \coordinate (#1_b3) at ($(#2)+(\cBx, {-\coneR*\isqrt}, \cBz)$);
  \coordinate (#1_b4) at ($(#2)+(\cBx, {\coneR*\isqrt}, \cBz)$);
  \pgfmathsetmacro{\cFx}{\farT*#3 + \farR*\isqrt*#5}
  \pgfmathsetmacro{\cFz}{\farT*#4 + \farR*\isqrt*#6}
  \coordinate (#1_f1) at ($(#2)+(\cFx, {\farR*\isqrt}, \cFz)$);
  \coordinate (#1_f2) at ($(#2)+(\cFx, {-\farR*\isqrt}, \cFz)$);
  \pgfmathsetmacro{\cFx}{\farT*#3 - \farR*\isqrt*#5}
  \pgfmathsetmacro{\cFz}{\farT*#4 - \farR*\isqrt*#6}
  \coordinate (#1_f3) at ($(#2)+(\cFx, {-\farR*\isqrt}, \cFz)$);
  \coordinate (#1_f4) at ($(#2)+(\cFx, {\farR*\isqrt}, \cFz)$);
}

\tikzfading[name=fadeout, inner color=transparent!10, outer color=transparent!90]
\newcommand{\drawcone}[3]{%
  \pgfmathsetmacro{\coneStep}{1/\coneLayers}
  \foreach \k in {1,...,\coneLayers} {
    \pgfmathsetmacro{\s}{\k*\coneStep}
    \fill[#3, opacity=\coneAlpha]
      (#2) -- ($(#2)!\s!(#1_f1)$) -- ($(#2)!\s!(#1_f2)$) -- cycle;
    \fill[#3, opacity=\coneAlpha]
      (#2) -- ($(#2)!\s!(#1_f2)$) -- ($(#2)!\s!(#1_f3)$) -- cycle;
    \fill[#3, opacity=\coneAlpha]
      (#2) -- ($(#2)!\s!(#1_f3)$) -- ($(#2)!\s!(#1_f4)$) -- cycle;
    \fill[#3, opacity=\coneAlpha]
      (#2) -- ($(#2)!\s!(#1_f4)$) -- ($(#2)!\s!(#1_f1)$) -- cycle;
  }
  \draw[#3, thin, opacity=0.3] (#2) -- (#1_f1) -- (#1_f2) -- cycle;
  \draw[#3, thin, opacity=0.3] (#2) -- (#1_f2) -- (#1_f3) -- cycle;
  \draw[#3, thin, opacity=0.3] (#2) -- (#1_f3) -- (#1_f4) -- cycle;
  \draw[#3, thin, opacity=0.3] (#2) -- (#1_f4) -- (#1_f1) -- cycle;
  \ifconeborders
    \draw[#3, opacity=0.5] (#1_b1) -- (#1_b2) -- (#1_b3) -- (#1_b4) -- cycle;
  \fi
}

\section{Introduction}
\looseness=-1
\IEEEPARstart{S}{ound} propagation in a real-world environment involves more than direct travel from sources to receivers. Numerous interactions with room surfaces and objects, e.g., reflections, scattering, and transmissions, result in early reflections and late reverberation, which characterize the environment's acoustic properties \cite{kuttruff2016room}.
Accurate and efficient modeling of such phenomena is essential for applications including architectural design \cite{aspock2014real}, gaming audio \cite{schissler2011gsound, cao2016interactive}, and augmented/virtual reality \cite{amengual2024perceptual, funkhouser1999real, potter2022relative}, yet remains a challenging research problem.

\looseness=-1
Recently, data-driven methods for learning sound propagation have emerged, commonly referred to as \emph{acoustic field learning} \cite{luo2022learning}. 
Given available observations and a (possibly approximate) room geometry, the goal is to predict acoustic measurements at novel source-receiver configurations.
This contrasts with the more established sound field estimation \cite{ueno2018kernel, miotello2024reconstruction, damiano2024compressive, ribeiro2024sound, antonello2017room, brunnstrom2025time, verburg2025differentiable}, which typically reconstructs a single pressure field within a local receiver region, though recent work has begun to relax this constraint \cite{figueroa2025reconstruction}.

Acoustic field learning has been explored using deep neural networks, considering different estimation targets: most commonly room impulse responses (RIRs) \cite{luo2022learning, su2022inras, majumder2022few, he2024deep, lan2024acoustic}, but also propagated speech \cite{chen2023novel, chen2024av, gao2024soaf} or acoustic parameters (e.g., reverberation time) \cite{falcon2024novel}.
These methods have reported promising results on both synthetic \cite{chen2020soundspaces} and real-world benchmarks \cite{chen2024real}.
However, they are typically trained on large-scale measurements from a single environment.
For example, the RIR predictors \cite{luo2022learning, su2022inras, majumder2022few, he2024deep, lan2024acoustic} have primarily been demonstrated as efficient codecs, compressing dense collections of RIRs to reduce storage usage.
Yet acoustic measurements are costly and time-consuming to acquire, and improving model performance under data-scarce conditions is desirable for practical deployment.

Several directions have been pursued to improve such data efficiency.
One approach is to formulate the task as few-shot learning: instead of training an independent neural network for each scene, a single model can be trained on multiple scenes to maximize data utilization and promote generalization \cite{majumder2022few, falcon2024novel, liu2025hearing}.
A different approach is to leverage auxiliary geometry-related information; extra visual input can be utilized to estimate room geometry \cite{chen2024av, gao2024soaf, liang2023av, brunetto2024neraf, bhosale2024av, liu2025hearing} or the room surface mesh can be directly used \cite{ratnarajah2022mesh2ir}.
Finally, some incorporate inductive biases by designing (or training) models to encode properties of sound propagation, such as the inverse square law \cite{lan2024acoustic}, acoustic reciprocity \cite{luo2022learning, he2024deep, lan2025resounding}, and propagation time delay \cite{lan2024acoustic}.
This brings the models closer to acoustic simulators.
However, it remains unclear whether these models combine such laws with room geometry information in a physically consistent manner.

\looseness=-1
On the other hand, \emph{differentiable signal processing} \cite{engel2020ddsp, hayes2024review} offers an alternative that directly incorporates the inductive biases of room acoustics.
By backpropagating through simulation models, one can optimize material properties to match the reference via gradient descent.
Various differentiable room acoustics models have been investigated, ranging from wave-based methods \cite{antonello2014identification} to geometric acoustics \cite{wang2024hearing, zhi2023differentiable, schissler2017acoustic, li2018scene, tang2020scene}. The former solve the wave equation \cite{kinsler2000fundamentals}, which is physically accurate but computationally expensive, limiting its use to small environments. The latter solve the acoustic rendering equation (ARE) \cite{siltanen2007room}, which simplifies sound propagation as rays, achieving efficiency at the cost of missing some wave effects, e.g., diffraction.
Remarkably, most prior differentiable geometric acoustics models \cite{antonello2014identification, wang2024hearing, zhi2023differentiable, schissler2017acoustic, li2018scene, tang2020scene} fixed material properties and optimized only reflection (or absorption) coefficients.
This overlooks the complex nature of sound propagation in a given room geometry and limits the model accuracy.
More recently, several works have integrated acoustic simulators with neural networks to compensate for inaccuracies or unmodeled effects \cite{jin2025avdar, liang2025p}.
While such hybrid approaches are promising, our focus is complementary: we aim to improve the expressiveness and learnability of differentiable geometric acoustics models themselves.

In this context, \emph{Acoustic Radiance Transfer} (ART) \cite{siltanen2007room, bai2013modeling, bai2015geometric, scerbo2024room, scerbo2024mod, siltanen2009frequency, siltanen2011efficient} appears to be a particularly attractive geometric acoustics model.
ART computes time- and direction-dependent energy exchange between surface patches through discretization, transforming the ARE into a finite linear system. This formulation enables the use of arbitrary material properties while retaining high runtime efficiency. Nevertheless, ART has not yet been considered within a differentiable signal processing framework.

To fill this gap, we propose DART, a differentiable implementation of ART.
The primary challenge of realizing DART is the high computational cost; ART achieves its noted efficiency only at runtime when its materials remain fixed, a condition that does not hold during optimization.
We overcome this challenge by introducing multiple techniques, including kernel factorization, frequency-domain formulations, and sparse storage/operations. As a result, we significantly reduce memory and computational cost while retaining flexible material parameterization.

We evaluate DART on a variant of acoustic field learning: we predict a time-energy response (called the \emph{echogram}) of a novel source-receiver pair from given observations, rather than the full RIR waveform.
While this task might appear simpler, our investigation reveals that it is still highly challenging in many practical scenarios, such as sparse measurements (from $4$ to $48$), multi-room scenes with diverse acoustic characteristics, and evaluation on unseen source/receiver regions.
Existing baselines, such as the differentiable image-source method \cite{wang2024hearing} and neural network-based approaches \cite{su2022inras, lan2024acoustic}, struggle in these settings, often failing to beat simple nearest-neighbor or linear interpolation. On the other hand, due to its grounded physical formulation, DART consistently shows strong performance.

We also analyze DART in depth. We explore and compare different material parameterizations and reveal the importance of flexible material properties.
We test DART's robustness to geometric distortion, validate design choices through ablation studies, and assess its computational and memory costs.
Finally, we share our implementation for room acoustics and machine learning research.
Notably, DART is the first open-source ART besides simplified versions \cite{interactiveacoustics, fatelaefficient}. 
\begin{figure}[t]
\centering
\tdplotsetmaincoords{\viewElev}{\viewAzim}
\begin{tikzpicture}[tdplot_main_coords, >=Stealth, semithick, scale=\viewscale]

\ifshowaxes
\foreach \x in {0,1,2,3,4,5} {
  \draw[gray!25, very thin] (\x, 0, 0) -- (\x, 4, 0);
}
\foreach \y in {0,1,2,3,4} {
  \draw[gray!25, very thin] (0, \y, 0) -- (5, \y, 0);
}
\foreach \y in {0,1,2,3,4} {
  \draw[gray!25, very thin] (0, \y, 0) -- (0, \y, 3);
}
\foreach \z in {0,1,2,3} {
  \draw[gray!25, very thin] (0, 0, \z) -- (0, 4, \z);
}
\draw[->, gray!60, thin] (0,0,0) -- (5.5, 0, 0) node[right, font=\small] {$x$};
\draw[->, gray!60, thin] (0,0,0) -- (0, 4.5, 0) node[above, font=\small] {$y$};
\draw[->, gray!60, thin] (0,0,0) -- (0, 0, 3.5) node[above, font=\small] {$z$};
\fi
\fill[\incomingcolor!15, draw=\incomingcolor, thick]
  (\chx, {\chy-\patchHY}, {\chz-\patchHZh}) -- (\chx, {\chy+\patchHY}, {\chz-\patchHZh})
  -- (\chx, {\chy+\patchHY}, {\chz+\patchHZh}) -- (\chx, {\chy-\patchHY}, {\chz+\patchHZh}) -- cycle;
\node[inner sep=0pt, \incomingcolor, left] at (\chx - 0.06, \chy, {\chz+0.9}) {$\A_h$};
\coordinate (ch) at (\chx, \chy, \chz);

\fill[gray!20, draw=black, thick]
  ({\cix-\patchHXi}, {\ciy-\patchHY}, \ciz) -- ({\cix+\patchHXi}, {\ciy-\patchHY}, \ciz)
  -- ({\cix+\patchHXi}, {\ciy+\patchHY}, \ciz) -- ({\cix-\patchHXi}, {\ciy+\patchHY}, \ciz) -- cycle;
\node[inner sep=0pt] at (\cix + .95, \ciy, {\ciz - 0.08}) {$\A_i$};
\coordinate (ci) at (\cix, \ciy, \ciz);

\computecone{hj}{ch}{\hjAx}{\hjAz}{\hjPx}{\hjPz}
\drawcone{hj}{ch}{\incomingcolor}
\node[\incomingcolor] at ($(ch)+(1.90, 0.55, -0.85)$) {$\S_{hj}$};

\computecone{ik}{ci}{\ikAx}{\ikAz}{\ikPx}{\ikPz}
\drawcone{ik}{ci}{\incidentcolor}
\node[\incidentcolor] at ($(ci)+(1, 0.55, 1.7)$) {$\S_{ik}$};

\draw[dashed, gray, thin] (ch) -- (ci);

\pgfmathsetmacro{\arrLen}{0.8}
\draw[->, thick] (ch) -- ++(\arrLen*\hjAx, 0, \arrLen*\hjAz)
  node[right, font=\small] {$v_{x\to x'}$};
\draw[->, thick] (ci) -- ++({-\arrLen*\hjAx}, 0, {-\arrLen*\hjAz})
  node[left, font=\small] {$v_{x'\to x}$};

\pgfmathsetmacro{\normLen}{0.8}
\draw[->, thick, gray] (ch) -- ++(\normLen, 0, 0)
  node[above, font=\small] {$\nu(x)$};
\draw[->, thick, gray] (ci) -- ++(0, 0, \normLen)
  node[above, font=\small] {$\nu(x')$};

\draw[->, thick] (ci) -- ++({0.9*\ikAx}, 0, {0.9*\ikAz});
\node[font=\small] at ($(ci)+({1.1*\ikAx}, 0, {1.1*\ikAz})$) {$\Omega$};

\fill[black] (ch) circle (1pt);
\node[left] at ($(ch)+(-0.02, 0, 0)$) {$x$};
\fill[black] (ci) circle (1pt);
\node[below] at ($(ci)+(0, 0, -0.01)$) {$x'$};

\end{tikzpicture}
\caption{
A diagram depicting the acoustic rendering equation (ARE) and acoustic radiance transfer (ART).
For a room boundary $\A$, ARE states that a radiance from a surface point $x'\in\A$ with outgoing direction $\Omega$ equals all other radiances heading to the point $x'$ properly attenuated and delayed with a reflection kernel $R$. ART discretizes the ARE into surface patches (two of which, $\A_h$ and $\A_i$, are depicted) and direction bins ($\S_{hj}$ and $\S_{ik}$, visualized as cones). Arrows depict the direction vectors used in the equations. Patches are shown as rectangles for clarity; in practice, triangle meshes are used.
}
\label{fig:art}
\end{figure}

\section{Background}\label{sec:background}
\subsection{Acoustic Rendering Equation}
\subsubsection{Definition}
Accounting for the non-negligible speed of sound, Siltanen et al. \cite{siltanen2007room} extended the rendering equation for light transport \cite{kajiya1986rendering} to the acoustic rendering equation (ARE).
Sound is regarded as a collection of ``rays,'' each carrying an acoustic radiance $L(x', \vout, t) \in \mathbb{R}^+$, defined as the time-varying energy flux per unit projected area per unit solid angle. It is a function of surface point $x' \in \calA$, emitting direction $\vout \in \calS$, and continuous time $t \in \mathbb{R}$, where $\A$ and $\S$ denote a room surface geometry and the set of all unit-length direction vectors, respectively. See Figure \ref{fig:art} for an illustration.
The ARE reads
\begin{equation}\label{eq:rare}    
\begin{split}
    L(x', \vout&, t) =  L^{(0)}(x', \vout, t) \\ 
    &  + \iint_\A R(x, x', \vout, t) * L(x, \vin, t) d\A(x).
\end{split}
\end{equation}
That is, the radiance is the sum of an initial term $L^{(0)}$ and the aggregation of other radiances convolved ($*$) with a reflection kernel $R$.
Here, $v_{x\to x'}=(x'-x)/\|x'-x\|$ denotes a unit-length direction vector from a surface point $x$ toward $x'$.
Formal definitions of $L^{(0)}$ and $R$ will be introduced later.
The use of radiance for room acoustics is valid when the characteristic dimensions of the room geometry are orders of magnitude larger than the longest wavelength of interest \cite{savioja2015overview}.

\subsubsection{Neumann Series}
The ARE has a so-called Neumann series solution \cite{kajiya1986rendering}, where the radiance $L$ can be obtained as an infinite sum of $\mathrm{k}$-th order terms,
\begin{equation} \label{eq:series-1}
    L(x', \vout, t) = L^{(0)}(x', \vout, t) + \sum_{k=1}^\infty L^{(\mathrm{k})}(x', \vout, t),
\end{equation}
where each $\mathrm{k}$-th order term $\smash{L^{(\mathrm{k})}}$ (for $\mathrm{k}\geq 1$) is obtained by recursive reflections of the initial radiance $\smash{L^{(0)}}$.
\begin{equation}\label{eq:series-2}
\begin{split}
    L^{(\mathrm{k}+1)}(&x', \vout, t) \\ &= \iint_\A R (x, x', \vout, t) * L^{(\mathrm{k})}(x, \vin, t) d\A(x).
\end{split}
\end{equation}
In practice, the series is truncated at a finite order, chosen large enough that the remaining energy is perceptually negligible.

\subsubsection{Reflection Kernel}
The reflection kernel is typically decomposed into four factors as
\begin{equation}\label{eq:reflection-kernel}
\begin{split}
    R(x, x', \vout, t) = V(x, x')  G(x, x')   D (x, x', t) \rho (v_{x'\to x}, \vout; x') .
\end{split}
\end{equation}
The first term, $V(x, x')$, is a visibility function that equals $1$ if $x$ and $x'$ are in an unobstructed line of sight, $0$ otherwise.
The second term is known as a patch-to-patch geometry, given as
\begin{equation}
    G(x, x') = \frac{|v_{x'\to x}\cdot \nu(x')||v_{x\to x'}\cdot\nu(x)|}{\|x-x'\|^2} 
\end{equation}
where $\nu(x)$ denotes the unit-length normal vector of the surface $\A$ at $x \in \A$ and each dot product computes the cosine of the angle between the two vectors.
Its numerator is the product of two absolute cosines, corresponding to Lambert's cosine law for the source and reflecting patches. Its denominator corresponds to the inverse square law. 
The third term models propagation delay, given as
\begin{equation}
    D(x, x', t) = \delta \bigg(t - \frac{\|x-x'\|}{c}\bigg)
\end{equation} 
where $\delta(\cdot)$ and $c$ denote the Dirac delta and the speed of sound, respectively. We omit the attenuation due to air absorption for simplicity.  
These three terms, visibility, patch-to-patch geometry, and delay, depend only on the room geometry $\calA$.

\looseness=-1
The last term $\rho(v_{x'\to x}, \Omega; x')$ is the bidirectional reflectance distribution function (BRDF) \cite{nicodemus1965directional}, which describes how incident energy arriving from direction $v_{x'\to x}$ is reflected into an outgoing direction $\Omega$ at the surface point $x'$.
In this work, we generalize this to the bidirectional scattering distribution function (BSDF) \cite{bartell1981theory}, which, to our knowledge, has not previously been explored in ART.
While the BRDF is defined over a single hemisphere, modeling either exterior or interior reflection, the BSDF extends both input and output domains to the full sphere $\S$.
This enables the modeling of transmission, which is common in complex multi-room geometries.
We nonetheless retain the term ``reflection kernel'' for consistency with the existing literature.

\subsubsection{Injection}
The injection process computes the initial radiance $L^{(0)}$, which represents the result of the emitted sound undergoing its first reflection (or transmission) at the room surface.
For a source $s$ emitting a unit impulse $\delta(t)$, the initial radiance is given by
\begin{equation}\label{eq:injection}
    \begin{split}
    L^{(0)}(x, \vin, t) = &\frac{\Gamma_s(v_{x_s\to x}, o_s)}{4\pi} \frac{|v_{x\to x_s}\cdot \nu(x)|}{\|x-x_s\|^2} \\ &\;\;\;V(x_s, x) D(x_s, x, t) 
    \rho (v_{x'\to x}, \vout; x')
    \end{split}
\end{equation}
where $x_s$, $o_s$, and $\Gamma_s$ denote source position, orientation, and directivity, respectively.

\subsubsection{Detection}
A receiver $r$ with position $x_r$, orientation $o_r$, and directivity $\Gamma_r$ receives the power from the radiances as
\begin{equation}\label{eq:detection}
\begin{split}
    E_L(t) =  \iint_{\A} &  L(x, v_{x\to x_r}, t) *  \Gamma_r(v_{x_r\to x}, o_r)  \\ & \frac{|v_{x\to x_r}\cdot \nu(x)|} {\|x-x_r\|^2}  V (x, x_r) D (x, x_r, t)   d\calA(x). 
\end{split}
\end{equation}
Finally, with the direct arrival from the source to the receiver,
\begin{equation}\label{eq:direct-arrival}
\begin{split}
    E_D(t) & = \frac{\Gamma_s(v_{x_s\to x_r}, o_s)\Gamma_r(v_{x_r\to x_s}, o_r)}{4\pi \|x_r - x_s\|^2} \\ & \quad \quad \quad \quad \quad \quad \quad \quad \quad \quad
    V(x_s, x_r) D(x_s, x_r, t),
\end{split}
\end{equation}
we obtain a time-varying response called the \emph{echogram}  
\begin{equation}
E(t) = E_D(t) + E_L(t),
\end{equation}
which describes energy transfer from the source $s$ to the receiver $r$ under the room geometry $\A$. 
Each echogram can be used to synthesize an RIR with the so-called auralization process, e.g., noise shaping \cite{kuttruff1993auralization}.

\begin{figure*}
    \centering
    \includegraphics[clip, trim={0mm .5mm 0mm 0mm}, width=.9\linewidth]{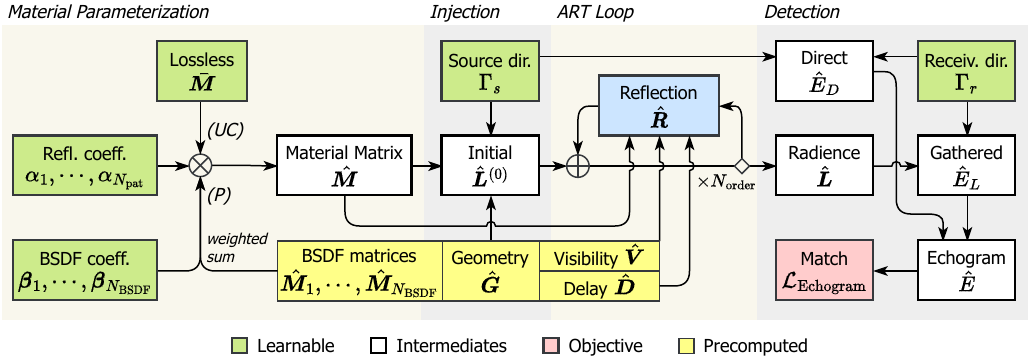}
    \caption{
    \looseness=-1
    Overview of DART.
    An echogram is computed in three stages: injection, iterative reflections (``ART loop''), and detection.
    The reflection, highlighted in a blue box, is decomposed into three components: delay, mean visibility, and material matrix. The first two can be precomputed, reducing the optimization cost. The third component, the material matrix, depends on material parameters and is therefore computed for every optimization step.
    The material matrix is also used for the injection stage.
    The material matrix has unconstrained (UC) and parametric (P) variants.  Learnable source or receiver directivities can optionally be introduced.
    All signal flows are differentiable, so that the echogram prediction loss can be backpropagated to all learnable components (shown as green boxes) end-to-end.
    }
    \label{fig:framework}
\end{figure*}

\subsection{Acoustic Radiance Transfer}
\subsubsection{Discretization} We first discretize the geometry into $\Npat$ patches.
The set of all incoming and outgoing directions (the unit sphere $\S$) is also partitioned into $N_\mathrm{dir}$ bins independently for each patch $\A_i$.
\begin{equation}\label{eq:discretization}
    \A = \A_1\cup \cdots\cup \A_{\Npat}, \quad \calS = \calS_{i 1}\cup \cdots \cup \calS_{iN_\mathrm{dir}}.
\end{equation}
Then, in the acoustic radiance transfer (ART) \cite{siltanen2007room}, a discrete radiance is defined as an average of the continuous radiance over the discretized patch $\A_i$ and direction bin $\S_{ik}$.
The continuous time $t$ is also discretized with sampling period $\Delta t$, yielding a discrete index $n$.
\begin{equation}\label{eq:discrete-radiance}
    \hat{L}_{i  k}[n] = \iint_{\calS_{i k}} \iint_{\A_i} L(x', \Omega, n\Delta t) \frac{d\A (x')}{|\A_i|} \frac{d\Omega}{|\calS_{i k}|}.
\end{equation}
Here, $|\A_i|$ and $|\S_{ik}|$ denote the area of the surface patch $\A_i$ and the solid angle of the direction bin $\S_{ik}$, respectively.

Assume that the continuous radiance is approximately constant within each discretized patch and direction bin:
\begin{equation}\label{eq:discrete-similar-assumption}
    L(x', \Omega, n \Delta t)\approx \hat{L}_{ik}[n], \quad x'\in \calA_i, \Omega \in \calS_{ik}.
\end{equation}
Then, for a discretized reflection kernel defined as
\begin{equation}\label{eq:discrete-reflection-kernel}
\begin{split}
\hat{R}_{h j, i k} [n] =  & \iint_{\calS_{i k}} \iint_{\A_i} \iint_{\calS_{h j}} \iint_{\A_h} \\ & \tilde{R}(x, \Theta, x', \Omega, n \Delta t)
 d\A (x) d\Theta \frac{d\A (x')}{|\A_i|} \frac{d\Omega}{|\calS_{i k}|}
\end{split}
\end{equation}
where we used the auxiliary reflection kernel given as 
\begin{equation}
    \tilde{R}(x, \Theta, x', \Omega, t) = R(x, x', \Omega, t) \cdot \delta (v_{x\to x'} - \Theta),
\end{equation}
we obtain the discrete version of the ARE.
\begin{equation}\label{eq:discrete-are}
\begin{split}
    \hat{L}_{i k}[n] 
    &\approx \hat{L}^{(0)}_{i k}[n] + \sum_{h=1}^{N_\mathrm{pat}} \sum_{j=1}^{N_\mathrm{dir}} \hat{R}_{h j, i k} [n] \circledast \hat{L}_{h j}[n]. \\
\end{split}
\end{equation}
Here, $\circledast$ denotes the discrete-time convolution. Throughout the paper, hatted symbols denote discretized quantities. 
Injection and detection can also be discretized accordingly, as we show later. The discrete ARE, injection, and detection constitute the ART system, which converges to its continuous counterpart as the discretization becomes finer.

\subsubsection{Delay Network Interpretation}
ART can be interpreted as belonging to the family of delay networks, filter structures that recursively interconnect multiple parallel delay lines.
We derive the explicit relationship in Appendix~\ref{app:delay-network}.
Also belonging to this family are widely used artificial reverberation algorithms, such as the feedback delay network (FDN) \cite{stautner1982designing, gerzon1971reverb} and scattering delay network (SDN) \cite{de2015efficient}.
However, ART differs in several aspects, including that it models energy rather than pressure and operates at a much larger scale.
As an example, consider a room geometry mesh with $N_\mathrm{pat} = 400$ surface patches and $N_\mathrm{dir} = 256$ direction bins.
The total number of parallel delay lines is given by the number of radiances, reaching roughly $N_\mathrm{rad} \approx 100\si{k}$, orders of magnitude more than the tens of delay lines typical of FDNs and SDNs.

\subsubsection{Frequency-domain ART}
Rather than solving the time-domain ARE \eqref{eq:discrete-are} directly, one can solve its frequency-domain counterpart to accelerate the radiance computation on a GPU.
Consider a stack of all discrete radiances in the $z$-domain;
we denote such a collection with a boldface letter:
\begin{equation}\label{eq:radiance-transfer-function}
\begin{split}
    \hat{\bm{L}}(z)  = [\hat{L}_{11}(z), \hat{L}_{12}(z),
    \cdots, \hat{L}_{N_\mathrm{pat}N_\mathrm{dir}}(z)]^T. \\ 
\end{split}
\end{equation}
Similarly, let $\smash{\hat{\bm{R}}}(z)$ denote the matrix collecting all discretized reflection kernel elements.
In this notation, the frequency-domain counterpart of the discrete ARE \eqref{eq:discrete-are} reads
\begin{equation}\label{eq:discrete-are-frequency}
\begin{split}
\hat{\bm{L}}(z) \approx \hat{\bm{L}}^{(0)}(z) + \hat{\bm{R}}(z) \hat{\bm{L}}(z), 
\end{split}
\end{equation}
and the corresponding Neumann series (\ref{eq:series-1} and \ref{eq:series-2}) can also be written as \cite{siltanen2007room}
\begin{subequations}\label{eq:series-frequency}
\begin{align}
    \hat{\bm{L}}(z) &\approx \hat{\bm{L}}^{(0)}(z) + \sum_{k=1}^\infty \hat{\bm{L}}^{(\mathrm{k})}(z), \\  
    \hat{\bm{L}}^{(\mathrm{k})}(z) &= \hat{\bm{R}}(z)\hat{\bm{L}}^{(\mathrm{k}-1)}(z).
\end{align}
\end{subequations}

Now, to obtain a length-$T$ echogram, we first solve the above $z$-domain ARE \eqref{eq:discrete-are-frequency} at the uniform samples on the upper half of the unit circle, i.e.,
\begin{equation}
    z = z_{T, f} = \exp \frac{2\pi \jmath f }{T}, \quad f =0, \cdots, \lfloor T/2 \rfloor,
\end{equation}
where $\jmath = \sqrt{-1}$.
Each solution can be obtained either by (i) direct inversion, i.e., explicitly forming the inverse matrix as in most prior differentiable delay networks \cite{lee2022differentiable, dal2024rir2fdn, dal2023differentiable}, (ii) direct solve, i.e., solving the linear system using factorization methods \cite{siltanen2011efficient}, or (iii) computing the Neumann series \eqref{eq:series-frequency} truncated to a finite order $N_\mathrm{order}$.
All three benefit from GPU parallelism, since each frequency bin has an independent linear system.

For DART, we choose the truncated Neumann series, since direct inversion is too costly at this scale.
The series approach also allows us to leverage the sparsity of the reflection kernel, significantly reducing the computation and memory costs (see the following section).
Explicit time-domain recursion \cite{mezza2024data, mezza2025differentiable} is also possible but requires $T \gg N_\mathrm{order}$ sequential steps, underutilizing GPU parallelism.

\newpage

After solving the frequency-domain systems, we apply the inverse Fast Fourier Transform (IFFT) to obtain time-domain signals that approximate the first $T$ samples of the discrete radiances $\smash{\hat{\bm{L}}[n]}$.
The source of error is time-aliasing, 
where the tails $n\geq T$ of the original signals are folded back into the range $0 \leq n<T$ \cite{smith2007mathematics}. 
Therefore, unless additional techniques are applied,
a sufficient number of samples $T$ is required to compute the echogram $\smash{\hat{E}[n]}$ accurately.

\section{Methods}\label{sec:methods}
We aim to optimize the material properties of the surface patches.
However, under the current ART formulation, every optimization step updates the material properties, requiring the discrete reflection kernel \eqref{eq:discrete-reflection-kernel} to be recomputed, which is prohibitively expensive.
Also, convolution with the kernel (\ref{eq:discrete-are} or \ref{eq:discrete-are-frequency}) is costly and memory-intensive.
With the example from Section \ref{sec:background} ($N_\mathrm{rad} \approx 100\si{k}$),
storing the dense reflection kernel alone would require roughly $\smash{N_\mathrm{rad}^2 \approx 40\si{GB}}$ of GPU memory per discrete time index $n$ in single floating point precision.
To address this challenge, we introduce several techniques that together minimize the optimization cost and make DART feasible. See Figure \ref{fig:framework} for an overview.

\subsection{Kernel Decomposition}
Our first key ingredient is a kernel decomposition that decouples the effects of the fixed room geometry from the learnable material properties, allowing us to precompute the former before optimization.

\subsubsection{Delay Decomposition}
First, following prior ART methods \cite{siltanen2007room, bai2015geometric, scerbo2024room}, we separate the propagation delay as
\begin{equation}\label{eq:kernel-decomposition}
    \hat{R}_{h j, i k}[n] \approx \hat{D}_{hj}[n] \cdot  \hat{S}_{hj,ik}.
\end{equation}
$\smash{\hat{D}_{hj}[n]}$ denotes a discrete delay signal for the radiance $\smash{\hat{L}_{hj}}[n]$ and $\smash{\hat{S}_{hj,ik}}$ is a time-independent, purely scalar-valued matrix.
Factoring out the time-dependent propagation delay significantly lowers the computational cost, reducing the number of convolutions from $\smash{N_\mathrm{rad}^2}$ to $N_\mathrm{rad}$.

Each discrete delay is realized with a linear fractional delay \cite{laakso1996splitting} to ensure nonnegativity of energy quantities, given as
\begin{equation}
    \hat{D}_{hj}[n] = a_{hj} \delta[n-\lfloor\hat{d}_{hj}\rfloor] + (1 - a_{hj}) \delta[n-\lceil\hat{d}_{hj}\rceil],
\end{equation}
where $\lfloor\cdot\rfloor$ and $\lceil\cdot\rceil$ denote floor and ceil operations, respectively.
Its linear weight and delay length are given as
\begin{subequations}
\begin{align}
    a_{hj} &= \lceil \hat{d}_{hj} \rceil - \hat{d}_{hj}, \\
    \hat{d}_{hj} &= \frac{\iint_{\calS_{h j}} \iint_{\A_h}  V_\A(x, \Theta) \|x - x'(x, \Theta)\| d\A (x) d\Theta}{c\Delta t \iint_{\calS_{h j}} \iint_{\A_h} V_\A(x, \Theta) d\A (x) d\Theta}. \label{eq:prop-delay}
\end{align}
\end{subequations}
Here, $V_\A(x, \Theta)$ is an indicator function that equals $1$ if the ray emitted from $x$ in direction $\Theta$ hits any point on the geometry $\A$.
$x'(x, \Theta) \in \A$ denotes a surface point where the ray first hits.
Hence, the delay length $\smash{\hat{d}_{hj}}$ is the average travel delay of the valid rays emitted from patch $\A_h$ within direction bin $\S_{hj}$ that hit any point on the geometry $\A$.

\subsubsection{Visibility-Material Decomposition}
After the above decomposition, the remaining scalar-valued matrix is given as
\begin{equation}\label{eq:scattering-kernel}
\begin{split}
& \hat{S}_{h j, i k} =  \iint_{\calS_{i k}} \iint_{\A_i} \iint_{\calS_{h j}} \iint_{\A_h} V(x, x')  G(x, x') \\ & \; \; \; \rho (v_{x'\to x}, \vout; x')  \delta (v_{x\to x'} - \Theta)  d\A (x) d\Theta \frac{d\A (x')}{|\A_i|} \frac{d\Omega}{|\calS_{i k}|},
\end{split}
\end{equation}
which still requires the nested integration of multiple terms, including the BSDF.
To reduce complexity further, we introduce an additional decomposition step (see Figure \ref{fig:art-decomposed} for illustration):
\begin{equation}\label{eq:scattering-decomposition}
\begin{split}
    \hat{S}_{hj,ik} \approx \sum_{l=1}^{\Ndir}{\hat{V}_{hj,il}}  {\hat{M}_{il,ik}}.
\end{split}
\end{equation}
The first term is the \emph{mean visibility matrix}, which quantifies the fraction of the propagated source radiance $\smash{\hat{D}_{hj}[n]*\hat{L}_{hj}}[n]$ visible from the receiving patch $\A_i$ with direction bin $\S_{il}$.
It is defined as
\begin{equation}\label{eq:mean-visibility-matrix}
\begin{split}
    \hat{V}_{hj,il} = {\iint_{\calS_{i   l}} \iint_{\A_i} V_{h   j}(x', \Phi) \frac{d\A (x')}{|\A_i|} \frac{d\Phi}{|\calS_{i   l}|}},
\end{split}
\end{equation}
where $V_{hj}(x', \Phi)$ is another indicator function that equals $1$ if the ray emitted from $x'$ in direction $\Phi$ will hit patch $\A_h$ with an incident direction falling within $\calS_{hj}$.
By aggregating all delayed source radiances attenuated by the mean visibility matrix, we obtain an incident radiance $\smash{\hat{L}^\mathrm{(in)}_{il}[n]}$ for each reflecting or transmitting patch $\A_i$ and incident direction $\S_{il}$.

\begin{figure}[t]
\centering
\tdplotsetmaincoords{\viewElev}{\viewAzim}
\begin{tikzpicture}[tdplot_main_coords, >=Stealth, semithick, scale=\viewscale]

\ifshowaxes
\foreach \x in {0,1,2,3,4,5} {
  \draw[gray!25, very thin] (\x, 0, 0) -- (\x, 4, 0);
}
\foreach \y in {0,1,2,3,4} {
  \draw[gray!25, very thin] (0, \y, 0) -- (5, \y, 0);
}
\foreach \y in {0,1,2,3,4} {
  \draw[gray!25, very thin] (0, \y, 0) -- (0, \y, 3);
}
\foreach \z in {0,1,2,3} {
  \draw[gray!25, very thin] (0, 0, \z) -- (0, 4, \z);
}
\draw[->, gray!60, thin] (0,0,0) -- (5.5, 0, 0) node[right, font=\small] {$x$};
\draw[->, gray!60, thin] (0,0,0) -- (0, 4.5, 0) node[above, font=\small] {$y$};
\draw[->, gray!60, thin] (0,0,0) -- (0, 0, 3.5) node[above, font=\small] {$z$};
\fi
\fill[\incomingcolor!15, draw=\incomingcolor, thick]
  (\chx, {\chy-\patchHY}, {\chz-\patchHZh}) -- (\chx, {\chy+\patchHY}, {\chz-\patchHZh})
  -- (\chx, {\chy+\patchHY}, {\chz+\patchHZh}) -- (\chx, {\chy-\patchHY}, {\chz+\patchHZh}) -- cycle;
\node[inner sep=0pt, \incomingcolor, left] at (\chx - 0.06, \chy, {\chz+0.9}) {$\A_h$};
\coordinate (ch) at (\chx, \chy, \chz);

\fill[gray!20, draw=black, thick]
  ({\cix-\patchHXi}, {\ciy-\patchHY}, \ciz) -- ({\cix+\patchHXi}, {\ciy-\patchHY}, \ciz)
  -- ({\cix+\patchHXi}, {\ciy+\patchHY}, \ciz) -- ({\cix-\patchHXi}, {\ciy+\patchHY}, \ciz) -- cycle;
\node[inner sep=0pt] at (\cix + .95, \ciy, {\ciz - 0.08}) {$\A_i$};
\coordinate (ci) at (\cix, \ciy, \ciz);

\computecone{hj}{ch}{\hjAx}{\hjAz}{\hjPx}{\hjPz}
\drawcone{hj}{ch}{\incomingcolor}

\computecone{il}{ci}{\ilAx}{\ilAz}{\ilPx}{\ilPz}
\drawcone{il}{ci}{\outgoingcolor}

\computecone{ik}{ci}{\ikAx}{\ikAz}{\ikPx}{\ikPz}
\drawcone{ik}{ci}{\incidentcolor}

\pgfmathsetmacro{\arrLen}{0.8}
\draw[gray, ->, thick] (ch) -- ++(\arrLen*\hjAx, 0, \arrLen*\hjAz);
\node[gray, font=\small] at ($(ch)+({.95*\hjAx}, 0, {.95*\hjAz})$) {$\Theta$};
\draw[gray, ->, thick] (ci) -- ++({-\arrLen*\hjAx}, 0, {-\arrLen*\hjAz});
\node[gray, font=\small] at ($(ci)+({-0.95*\hjAx}, 0, {-0.95*\hjAz})$) {$\Phi$};
\draw[gray, ->, thick] (ci) -- ++({0.9*\ikAx}, 0, {0.9*\ikAz});
\node[gray, font=\small] at ($(ci)+({1.1*\ikAx}, 0, {1.1*\ikAz})$) {$\Omega$};

\coordinate (Lhj) at ($(ch)+(1.45, 0, 0.15)$);
\coordinate (Lil) at ($(ci)+(-0.15, 0, 1.34)$);
\node[\incomingcolor, left] (LhjLabel) at (Lhj) {$\hat{L}_{hj}[n]$};
\node[\outgoingcolor] (LilLabel) at (Lil) {$\hat{L}^{\mathrm{(in)}}_{il}[n]$};
\draw[->, thick, darkgray] (LhjLabel) to[bend left=17]
  node[pos=0.25, above right=-3pt, font=\small] (Dnode) {$*\hat{D}_{hj}[n]$}
  node[pos=0.75, above right=-3pt, font=\small] (Vnode) {$\times \hat{V}_{hj,il}$}
  (LilLabel);

\coordinate (Lik) at ($(ci)+(1.9, -0.3, 2.5)$);
\node[\incidentcolor] (LikLabel) at (Lik) {$\hat{L}_{ik}[n]$};
\draw[->, thick, darkgray] (LilLabel) to[bend left=20]
  node[pos=0.6, above=2pt, font=\small] (Mnode) {$\times \hat{M}_{il,ik}$}
  (LikLabel);

\begin{scope}[tdplot_screen_coords]
  \coordinate (midVM) at ($(Vnode)!0.5!(Mnode)$);
  \node[darkgray, font=\small, fill=white, inner sep=4pt] (Snode) at ($(midVM)+(0,0.85)$) {$\times\hat{S}_{hj,ik}$};
  \draw[darkgray, thin] (Snode) -- (Vnode);
  \draw[darkgray, thin] (Snode) -- (Mnode);
  \node[darkgray, font=\scriptsize, fill=white, inner sep=0pt] at ($(Snode)!0.5!(midVM)$) {\eqref{eq:scattering-decomposition}};
  \coordinate (midDS) at ($(Dnode)!0.5!(Snode)$);
  \node[darkgray, font=\small, fill=white, inner sep=4pt] (Rnode) at ($(midDS)+(0,0.85)$) {$*\hat{R}_{hj,ik}[n]$};
  \draw[darkgray, thin] (Rnode) -- (Dnode);
  \draw[darkgray, thin] (Rnode) -- (Snode);
  \node[darkgray, font=\scriptsize, fill=white, inner sep=0pt] at ($(Rnode)!0.5!(midDS)+(0.04,0)$) {\eqref{eq:kernel-decomposition}};
\end{scope}

\end{tikzpicture}
\caption{
{Decomposition of the reflection kernel $\smash{\hat{R}_{hj,ik}}[n]$ and its interpretation}. The tree shows the two-level factorization into the propagation delay $\smash{\hat{D}_{hj}[n]}$, mean visibility matrix $\smash{\hat{V}_{hj,il}}$, and material matrix $\smash{\hat{M}_{il,ik}}$.
The first step \eqref{eq:kernel-decomposition} separates the propagation delay from the reflection kernel, so that it is applied to each source radiance $\smash{\hat{L}_{hj}[n]}$.
The second step \eqref{eq:scattering-decomposition} factorizes the remaining matrix $\smash{\hat{S}_{hj,ik}}$ into two matrices; each delayed radiance is first attenuated by the mean visibility matrix to yield the incident radiance $\smash{\hat{L}^{\mathrm{(in)}}_{il}[n]}$, and then multiplied by the material matrix to yield the outgoing radiance $\smash{\hat{L}_{ik}[n]}$.
The direction bins of source, incident, and outgoing radiance are visualized in shades of orange, green, and blue, respectively, and arrows $\Theta \in \S_{hj}$, $\Phi \in \S_{il}$, $\Omega \in \S_{ik}$ mark the corresponding directions as used in the text.
}
\label{fig:art-decomposed}
\end{figure}

The second term, the material matrix, models the effect of the BSDF and converts the incident radiance $\smash{\hat{L}^\mathrm{(in)}_{il}[n]}$ into contributions to the outgoing radiance $\smash{\hat{L}_{ik}}[n]$. It is defined as
\begin{equation}\label{eq:material-matrix}
\begin{split}
\hat{M}_{il,ik} = {\iint_{\calS_{i   k}} \iint_{\calS_{i   l}} \rho_i(\Phi, \Omega) |\Phi\cdot\nu_i| d\Phi \frac{d\Omega}{|\calS_{i   k}|}}.
\end{split}
\end{equation}
As in the ARE discretization \eqref{eq:discrete-similar-assumption}, both terms rely on a locally constant assumption: the BSDF $\rho_i$ is uniform within each patch $\A_i$, and the mean visibility is constant within each directional bin. The same assumptions underlie the injection approximation introduced later \eqref{eq:injection-wo-monte-carlo}; all such errors vanish as the discretization becomes finer. Refer to Appendix~\ref{app:kernel-decomposition} for the derivation.

The discrete delays $\smash{\hat{\bm{D}}[n]}$ and the mean visibility matrix  $\smash{\hat{\bm{V}}}$ can be computed once and stored prior to optimization, as they only depend on the fixed room geometry.

\subsection{Material Parameterization}
The material matrix $\smash{\hat{\bm{M}}}$ still requires numerically integrating the BSDFs during optimization.
We explore two strategies that sidestep this cost, corresponding to two variants of DART.

\subsubsection{Unconstrained (UC) Variant}
We can bypass the integration and directly learn the matrix entries, factorized into a reflection coefficient $\alpha_i \in [0, 1)$ per patch $\A_i$ and a lossless (energy-preserving) \cite{scerbo2024mod} matrix $\smash{\bar{\bm{M}}}$.
\begin{equation}
    \hat{M}_{il, ik}^\mathrm{(UC)} = \alpha_i \smash{\bar{M}_{il, ik}}, \quad \sum_{l=1}^{N_\mathrm{dir}}\bar{M}_{il,ik} = 1.
\end{equation}
The unconstrained approach is particularly useful when only approximate meshes are available, as is common in practice.
Geometric errors in such meshes, whether missing or misplaced fine details, produce rich diffraction and scattering patterns that can be approximated with complex BSDFs on the simplified geometry \cite{walter2007microfacet, kuttruff2016room}. The maximal flexibility of the material matrix allows the unconstrained DART to learn such details.

\subsubsection{Parametric (P) Variant}
Alternatively, we can model each patch's BSDF $\rho_i$ as a convex combination of predefined $N_\mathrm{BSDF}$ BSDFs, each denoted as $\rho_{i, m}$, as follows,
\begin{equation}
    \rho_i(\Phi, \Omega) = \alpha_i \sum_{m} \beta_{i, m} \rho_{i, m}(\Phi, \Omega), \quad \sum_{m=1}^{N_\mathrm{BSDF}}\beta_{i, m} =1,
\end{equation}
where $\beta_{i, m} \in [0, 1)$ denotes the combination weights.
Then, the corresponding material matrix also becomes a convex combination of the BSDF matrices.
\begin{equation}
    \hat{M}_{il, ik}^\mathrm{(P)} = \alpha_i \sum_{m=1}^{N_\mathrm{BSDF}} \beta_{i,m} \hat{M}_{il,ik, m}.
\end{equation}
Each BSDF matrix follows the definition of the material matrix \eqref{eq:material-matrix} and is given as
\begin{equation}
    \hat{M}_{il,ik,m} = \iint_{\calS_{i   k}} \iint_{\calS_{i   l}} \rho_{i, m}(\Phi, \Omega) |\Phi\cdot\nu_i| d\Phi \frac{d\Omega}{|\calS_{i   k}|},
\end{equation}
and it can be precomputed since each component BSDF $\rho_{i, m}$ is fixed.
We let both reflection coefficients $\alpha_i$ and combination weights $\beta_{i,m}$ be learnable.
As component BSDFs, we use the ideal specular and diffuse reflections \cite{siltanen2007room}.
For multi-room scenes, we also include specular and diffuse transmissions.
Their definitions are given as
\begin{subequations}
\begin{align}
\rho^{\mathrm{(DiffuseReflection)}}_i(\Phi, \Omega) &=
\frac{1}{\pi} \mathbbm{1}_{\mathrm{sgn}(\Phi \cdot \nu_i) = \mathrm{sgn}(\Omega \cdot \nu_i)}, \\
\rho^{\mathrm{(DiffuseTransmission)}}_i(\Phi, \Omega) &=
\frac{1}{\pi} \mathbbm{1}_{\mathrm{sgn}(\Phi \cdot \nu_i) \neq \mathrm{sgn}(\Omega \cdot \nu_i)}, \\
\rho^{\mathrm{(SpecularReflection)}}_i(\Phi, \Omega) &=
\frac{\delta(\Phi + \Omega - 2\nu_i)}{|\Phi \cdot \nu_i|}, \\
\rho^{\mathrm{(SpecularTransmission)}}_i(\Phi, \Omega) &= \frac{\delta(\Phi + \Omega)}{|\Phi \cdot \nu_i|}.
\end{align}
\end{subequations}
$\mathrm{sgn}(\cdot)$ denotes the sign function and $\mathbbm{1}$ is an indicator function, which equals $1$ when the condition holds, $0$ otherwise.
\looseness=-1
This parametric DART is consistent with standard material choices in room acoustics modeling, providing more compact and interpretable parameters ($\smash{N}_\mathrm{BSDF} + 1$ parameters per patch, versus $\smash{N_\mathrm{dir}^2} + 1$ for unconstrained DART).

\subsection{Radiance Computation} \label{subsec:radiance-computation}
\subsubsection{Aliasing Suppression}
For optimization, it is unnecessary to compute the full-length echogram because, after sufficient reflections, the sound field becomes diffuse, and the remaining response can be well approximated by an exponential decay.
However, using a short echogram length $T$ in the frequency domain leads to coarse sampling and introduces the aforementioned time-aliasing.
To mitigate this, we employ the recently proposed technique \cite{dal2025flamo}, sampling {\emph{outside}} the unit circle:
\begin{equation}\label{eq:aliasing-suppression}
\begin{split}
    z = z_{T, \gamma, f} = \gamma^{-1/T} \exp \frac{2\pi \jmath f }{T}, \quad f =0, \cdots, \lfloor T/2 \rfloor.
\end{split}
\end{equation}
Here, $\gamma \in (0, 1)$ denotes a constant circle radius that controls the amount of aliasing suppression.
By applying this sampling scheme to the truncated series \eqref{eq:series-frequency} with the decomposed kernel (\ref{eq:kernel-decomposition} and \ref{eq:scattering-decomposition}), we can in practice compute the radiance by
\begin{equation}\label{eq:series-freq-sampled}
    \hat{\bm{L}}[n] \approx \gamma^{-n/T} \cdot \mathrm{IFFT} \Bigg[\sum_{\mathrm{k}=0}^{N_\mathrm{order}}\hat{\bm{L}}^{(\mathrm{k})}(z_{T, \gamma, f})\Bigg], \\
\end{equation}
where the $\mathrm{k}$-th term is given as
\begin{equation}\label{eq:series-freq-sampled-update}
    \hat{\bm{L}}^{(\mathrm{k})}(z_{T, \gamma, f}) = \hat{\bm{M}} \hat{\bm{V}}
    \Big[
    \hat{\bm{D}}(z_{T, \gamma, f}) \odot \hat{\bm{L}}^{(\mathrm{k}-1)}(z_{T, \gamma, f})\Big]
\end{equation}
where $\odot$ denotes elementwise multiplication.
The frequency-domain initial radiances and delays can be obtained by
\begin{subequations}
\begin{align}
    \hat{\bm{L}}^{(0)}(z_{T, \gamma, f}) &= \mathrm{FFT} \Big[ \gamma^{n/T} \hat{\bm{L}}^{(0)}[n] \Big], \\
    \hat{\bm{D}}(z_{T, \gamma, f}) &= \mathrm{FFT} \Big[ \gamma^{n/T} \hat{\bm{D}}[n] \Big].
\end{align}
\end{subequations}
Sampling outside the unit circle damps the problematic tails $n\geq T$ with an exponentially decreasing window $\smash{\gamma^{n/T}}$, thereby reducing the aliasing by more than a factor of $\gamma$. 
We leave the derivation of this error bound in Appendix \ref{app:aliasing}.

\subsubsection{Leveraging Sparsity} \label{subsec:sparsity}
We implement the core computations using sparse tensors and operations.
For instance, prior to optimization, we identify discrete radiances that never send or receive energy and therefore never contribute to the echogram.
We can remove corresponding columns and rows from the tensors, which reduces the number of discrete radiances to $\smash{\bar{N}_\mathrm{rad}} < N_\mathrm{rad}$ and yields more compact tensors (denoted with a subscript $\mathrm{r}$).
Furthermore, even after the reduction, the mean visibility and material matrices are highly sparse; their numbers of nonzeros are denoted as $\smash{\bar{N}_\mathrm{vis}, \bar{N}_\mathrm{mat} \ll \bar{N}_\mathrm{rad}^2}$, respectively.
Hence, we compute the Neumann series terms \eqref{eq:series-freq-sampled-update} with sparse matrix multiplications $\mathrm{spmm}$ as
\begin{equation}\label{eq:sparse-computation}
    \hat{\bm{L}}^{(\mathrm{k})}_\mathrm{r} = \mathrm{spmm}(\hat{\bm{M}}_\mathrm{r}, \mathrm{spmm}(\hat{\bm{{V}}}_\mathrm{r}, \hat{\bm{D}}_\mathrm{r}  \odot \hat{\bm{L}}^{(\mathrm{k}-1)}_\mathrm{r})),
\end{equation}
where the tensor shapes are given as
\begin{equation}\label{eq:sparse-size}
    \hat{\bm{L}}^{(\mathrm{k})}_\mathrm{r}, \hat{\bm{L}}^{(\mathrm{k}-1)}_\mathrm{r}, \hat{\bm{D}}_\mathrm{r} \in \mathbb{C}^{\bar{N}_\mathrm{rad}\times \lfloor T/2 + 1\rfloor}, \quad \hat{\bm{M}}_\mathrm{r}, \hat{\bm{V}}_\mathrm{r} \in \mathbb{R}^{\bar{N}_\mathrm{rad}\times \bar{N}_\mathrm{rad}}.
\end{equation}
We use two consecutive sparse matrix multiplications because their product would be substantially denser than either factor.

\subsubsection{Complexity Analysis}
We applied factorization and sparse operations to the single dense reflection kernel.
As a result, the computational and memory complexities of a single reflection step in the frequency domain are improved, respectively, to:
\begin{subequations}
\begin{align}
    \mathcal{O}({N}_\mathrm{rad}^2 T) &\to \mathcal{O}(\bar{N}_\mathrm{rad}  T ) + \mathcal{O}(\bar{N}_\mathrm{vis} T) + \mathcal{O}(\bar{N}_\mathrm{mat} T),  \\
    \mathcal{O}({N}_\mathrm{rad}^2 T) &\to \mathcal{O}(\bar{N}_\mathrm{rad}T) + \mathcal{O}(\bar{N}_\mathrm{vis}) + \mathcal{O}(\bar{N}_\mathrm{mat}),
\end{align}
\end{subequations}
where the three terms correspond to the complexities of applying delays, mean visibility, and material matrices, respectively.
The actual sparsity ratios depend on the room geometry and discretization resolution; however, in practice, a significant fraction of discrete radiances can be pruned, and both matrices are highly sparse relative to $\bar{N}_\mathrm{rad}^2$, with densities on the order of $10^{-4}$ to $10^{-3}$. The visibility matrix is typically the sparser of the two by roughly an order of magnitude, since most patch-direction pairs lack line-of-sight connectivity.

\subsection{Injection and Detection}
\subsubsection{Injection}
We obtain the discrete initial radiance $\smash{\hat{L}_{ik}^{(0)}}[n]$ by discretizing its continuous counterpart \eqref{eq:injection}. Using analogous locally constant approximations (refer to Appendix \ref{app:injection} for the derivation), we can decompose it as
\begin{equation} \label{eq:injection-wo-monte-carlo}
\hat{L}^{(0)}_{i k}[n] \approx
 \sum_{l=1}^\Ndir {\hat{M}_{il,ik}}\frac {\hat{P}^{\mathrm{(in)}}_{il}[n]}{\hat{G}_{il}}
\end{equation}
where $\smash{\hat{P}^{\mathrm{(in)}}_{il}[n]}$ is the incident power from the source arriving at patch $\A_i$ from direction bin $\S_{il}$, defined as
\begin{equation}
    \hat{P}^{\mathrm{(in)}}_{il}[n] = \hat{D}_{s, il} [n] \iint_{\S} {\Gamma_s(\Theta, o_s)} V_{il}(x_s, \Theta) \frac{d\Theta}{4\pi}.
\end{equation}
Here, $\smash{\hat{D}_{s, il}[n]}$ is a discrete delay signal encoding the average propagation delay from the source $s$ to patch $\A_i$ in direction bin $\S_{il}$, and the integral accumulates the directivity-weighted visibility over all ray directions.
In practice, the incident power is computed via Monte Carlo, by shooting rays from the source and aggregating their contributions into the corresponding patch-direction bins.
The dividing term $\smash{\hat{G}_{il}}$ is a precomputable constant, called the integrated geometry,
which converts the incident power into incident radiance.
\begin{equation}
    {\hat{G}_{il} = |\calA_i| \iint_{\calS_{il}} |\Phi \cdot \nu_i| d\Phi }.
\end{equation}
The material matrix $\smash{\hat{M}_{il,ik}}$ then scatters the incident radiance into the initial (outgoing) radiance.

\looseness=-1
We apply the material matrix rather than an explicit BSDF evaluation for the first reflection, departing from prior ART methods. In those methods, the BSDF is sampled directly via ray tracing during injection; during optimization, this would require maintaining the BSDF explicitly and recomputing the material matrix \eqref{eq:material-matrix} from it at every step to keep the system consistent. By using the material matrix directly for injection, we avoid this recomputation. This introduces a minor approximation but also reduces variance from ray-tracing stochasticity, making DART less sensitive to the number of injection rays.

\subsubsection{Detection}
Similarly, discretizing the continuous detection \eqref{eq:detection} yields
\begin{equation} \label{eq:detection-main}
\hat{E}_L[n]  \approx  \sum_{i=1}^{\Npat}\sum_{k=1}^\Ndir
\hat{W}_{r, ik}[n] \circledast \hat{L}_{ik}[n]
\end{equation}
where $\smash{\hat{W}_{r, ik}[n]}$ is the detection weight for patch $\A_i$ and direction bin $\S_{ik}$, defined as
\begin{equation} \label{eq:detection-operator}
    \hat{W}_{r, ik}[n] = \hat{D}_{r, ik}[n] \iint_{\S}   \Gamma_r(\Theta, o_r) V_{ik}(x_r, \Theta)   d\Theta.
\end{equation}
As in the injection, $\smash{\hat{D}_{r, ik}[n]}$ encodes the propagation delay from patch $\A_i$ to the receiver $r$, and the integral accumulates the directivity-weighted visibility. The detection weight is convolved with the radiance $\smash{\hat{L}_{ik}[n]}$ and summed over all patch-direction pairs to yield the detected power $\smash{\hat{E}_L[n]}$.
In practice, we use the sparse radiance discussed in Section \ref{subsec:sparsity} for both injection and detection, which we omit for brevity.
Finally, after computing the direct arrival $\smash{\hat{E}_D[n]}$, we obtain the echogram $\smash{\hat{E}[n]}$, which is multiplied by an auxiliary learnable gain $g \in \mathbb{R}^+$ to match the scale of the target echograms.

\subsection{Learnable Directivity}\label{app:directivity}
In practical scenarios, the source directivity $\Gamma_s$ (or receiver directivity $\Gamma_r$) may be unknown. We employ a learnable, axially symmetric directivity module inspired by DiffRIR \cite{wang2024hearing}.
We introduce $K$ learnable parameters $\kappa_1, \cdots, \kappa_K$ at equispaced angles $\phi_k$ from $0$ to $\pi$.
For a ray with outgoing direction $\Theta$, let $\theta$ be its angle from the principal axis $o$. The directivity is a softmax-weighted sum of $\kappa_k$, with weights determined by the angular distance $|\theta - \phi_k|$ and a temperature $\lambda > 0$:
\begin{equation}
\Gamma (\Theta, o) = \sum_{k=1}^K \frac{\exp (-\lambda  |\theta - \phi_k|)}{\sum_{l=1}^K \exp (-\lambda |\theta - \phi_l|)} \kappa_k.
\end{equation}

\section{Experimental Setup}\label{sec:setup}
\subsection{Benchmarks}
We evaluate DART with two real-world datasets.
First, we use the Hearing Anything Anywhere (HAA) dataset \cite{wang2024hearing}.
The HAA dataset comprises four scenes: \texttt{Classroom}, \texttt{Hallway}, \texttt{Dampened}, and \texttt{Complex}.
Each scene features several hundred RIR measurements with varying receiver positions, all from a single fixed source.
We follow the same split as initially proposed: $12$ RIRs for training, half of the remaining RIRs for validation, and the remaining ones for testing.
The HAA dataset also provides rough room geometry meshes, which are used for DART and other baselines.

\begin{figure}[t]
  \centering
  \includegraphics[clip, trim={0cm -0.5cm 0cm -0.5cm}, width=.97\linewidth]{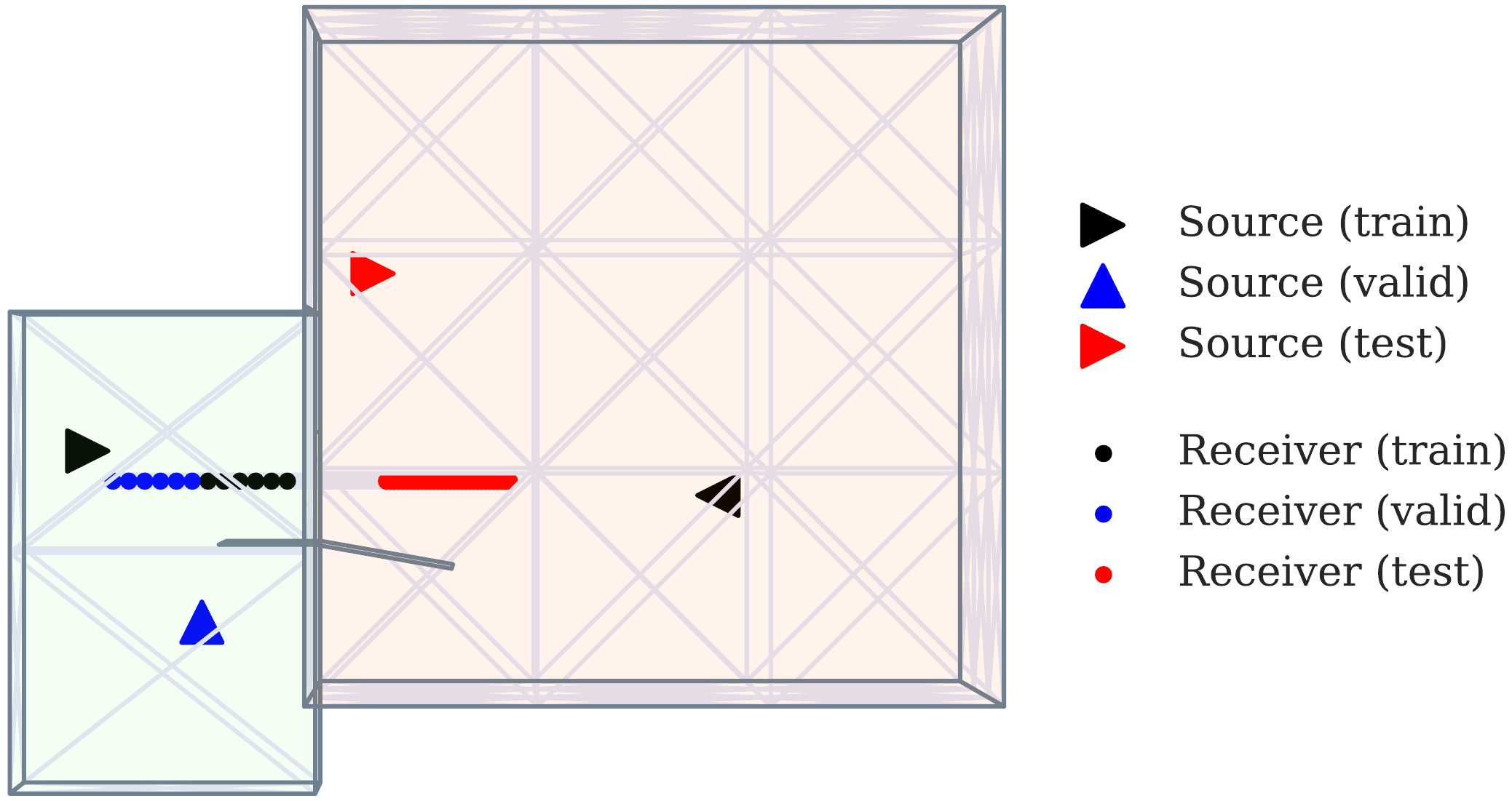} \\
  \caption{A top-down view of the \texttt{Office\;$\to$\;Anechoic} scene in the CR dataset with the unseen source/receiver split.
  \texttt{RoomA} and \texttt{RoomB} are shaded in green and red, respectively.
  The orientations of the sources are indicated by the triangles' orientations; the training and validation sources point towards the receivers, while the test source points towards the east.
  The triangle patches with thin lines are those produced by the patch subdivision procedure.
}
  \label{fig:cr-split}
\end{figure}

\looseness=-1
While the HAA dataset serves as an excellent benchmark, it also has some weaknesses: each scene consists of only one room, and is measured only with a single fixed source position and orientation.
This motivates us to complement it with a more challenging Coupled Room (CR) dataset \cite{mckenzie2021acoustic}.
The CR dataset also comprises four scenes: $\texttt{Office}\!\to\!\texttt{Anechoic}$, $\texttt{Office}\!\to\!\texttt{Stairwell}$, $\texttt{MeetingRoom}\!\to\!\texttt{Hallway}$, and $\texttt{Office}\!\to\!\texttt{Kitchen}$.
Each scene, denoted in the form \texttt{RoomA}\;$\to\;$\texttt{RoomB}, has two rooms, \texttt{RoomA} and \texttt{RoomB}.
$101$ receivers are positioned around the transition area from \texttt{RoomA} to \texttt{RoomB}, and four sources are considered with different positions and orientations, two in each room.
See Figure \ref{fig:cr-split} for an example; also refer to Figure \ref{fig:qualitative-unseen} and the supplementary for visualization of the remaining scenes.
Since the CR dataset provides a floor plan and the ceiling height for each scene, we construct each room geometry mesh ourselves. Similar to the HAA dataset, these meshes are coarser than the actual room geometries, omitting interior objects and fine architectural detail.

For the CR dataset, we consider two data splits to analyze the models' behavior under different scenarios.
In the first one, called \emph{random} split, we sample three receiver positions for each source (hence, a total of $12$ samples) and use them for training.
We sample another $12$ non-overlapping source-receiver pairs for validation in the same way, and the remaining samples form a test set.
The other one, called \emph{unseen} split, tests the models under completely unseen source/receiver pairs.
As shown in Figure \ref{fig:cr-split}, the training uses two sources, each with six receiver positions (again, a total of $12$ samples).
The validation uses one remaining source in \texttt{RoomA} with six different receivers.
For the testing, we use the remaining source in \texttt{RoomB} with $32$ different receivers.
Hence, this split tests whether the models can generalize and accurately estimate sound propagation, beyond merely interpolating the observations.

\looseness=-1
All RIRs are resampled to a $16\si{kHz}$ sampling rate.
The sampling rate of the discrete radiance is set to $1\si{kHz}$, as it has been shown to be sufficiently close to the continuous-time ART \cite{scerbo2024mod}.
We obtain each echogram by squaring the corresponding RIR and summing blocks of $16$ consecutive samples.
We consider RIRs or echograms of $0.32\si{s}$ length, i.e., $T=320$.

\subsection{Metrics}
We compare the predicted echograms and their corresponding ground truths with the sample-level $l_1$ distance (L1).
We also consider three reverberation parameter distances: reverberation time (T60, in \%), early decay time (EDT, in seconds), and clarity (C50, in dB).
Each model is validated at $25$ regular intervals during its optimization, and then the best checkpoint is selected for the final test.

\subsection{Implementation Details}
\subsubsection{Discretization}
For surface discretization, we subdivide the room geometry into small triangle patches. Specifically, we subdivide the raw parallelogram mesh into approximately equal-area parallelograms, ensuring no edge exceeds a specified maximum length, and then split each into four triangles along its diagonals.
The resulting patch count ranges from $120$ (\texttt{Hallway}, HAA) to $260$ ($\texttt{Office}\!\to\!\texttt{Anechoic}$, CR).
We let each subpatch have its own independent material parameters.
Each patch's direction sphere $\S$ is partitioned into a grid of $N_\mathrm{azi}=12$ azimuths and $N_\mathrm{ele}=12$ elevations, i.e., $N_\mathrm{dir} = 144$, with equal solid angles.

\subsubsection{Radiance Computation}
We set the truncated Neumann series order $N_\mathrm{order}$ for each scene based on its geometric scale.
We divide the total propagation length ($\approx 110\si{m}$ for a $0.32\si{s}$ echogram) by the shortest room dimension to upper-bound the number of repeated reflections, which sets the truncation order.
The resulting values range from $25$ for the \texttt{Complex} scene to $72$ for the \texttt{Hallway} scene.
We empirically observed that this choice was sufficient to model the dominant energy contributions within the echogram duration; increasing the order beyond these values produced only negligible changes in the predicted echograms.
For the time-aliasing suppression, we use $\gamma = 10^{-2}$, which reduces the aliasing error by more than $20\si{dB}$. We found this value to be sufficient, and using a smaller $\gamma$ amplified numerical noise due to the exponentially increasing window \eqref{eq:series-freq-sampled}, consistent with prior observations \cite{dal2025flamo}.
\subsubsection{Ray Tracing Monte Carlo}
Several ART components are computed via Monte Carlo ray tracing.
For the mean visibility matrix, we sample $10 \times 10$ points on each patch and shoot $4096$ rays from each point, totaling $N_\mathrm{precompute} = 100 \times 4096 \times N_\mathrm{pat}$ rays; the propagation delays are obtained from the same traces.
For injection and detection, we shoot $10\si{k}$ rays each, using stratified sampling on the patch and direction grid.

\subsubsection{Directivity}
As both datasets employ directional speakers and omnidirectional microphones, we learn only the source directivity $\Gamma_s$, with scale $\lambda = 8$ and $K=128$ directions, and omit the receiver directivity $\Gamma_r$.

\subsubsection{Core Libraries}
We use \texttt{PyTorch} \cite{paszke2019pytorch} for the implementation, with \texttt{pytorch\_sparse}\footnote{\tt \url{github.com/rusty1s/pytorch_sparse}} and \texttt{pytorch\_scatter}\footnote{\tt \url{github.com/rusty1s/pytorch_scatter}} for efficient sparse computation and backpropagation. \texttt{NVIDIA Warp} \cite{warp2022} is used for efficient ray tracing.

\subsection{Optimization}
\subsubsection{Learning Objective}
As a main optimization objective, we use an echogram loss defined as
\begin{equation}
    \calL_\mathrm{Echogram} = \calL_\mathrm{NMSE} + \calL_\mathrm{EDC}.
\end{equation}
$\calL_\mathrm{NMSE}$ is a normalized mean square error (NMSE) between the predicted echogram $\smash{\hat{E}^{\mathrm{pred}}[n]}$ and its ground-truth $\smash{\hat{E}^{\mathrm{true}}[n]}$.
\begin{equation}
\calL_\mathrm{NMSE} = \frac{\left\| \hat{E}^\mathrm{pred}[n] - \hat{E}^{\mathrm{true}}[n] \right\|^2}{{\left\|\hat{E}^{\mathrm{true}}[n]\right\|^2}}.
\end{equation}
$\calL_\mathrm{EDC}$ is a normalized mean absolute error of the energy decay curve (EDC), an inverse cumulative sum of an echogram \cite{jot1992analysis}.
\begin{equation}
    \calL_\mathrm{EDC} = \frac{\left\| \mathrm{EDC}(\hat{E}^\mathrm{pred})[n] - \mathrm{EDC}(\hat{E}^{\mathrm{true}})[n] \right\|_1}{\left\|\mathrm{EDC}(\hat{E}^{\mathrm{true}})[n]\right\|_1}.
\end{equation}
The NMSE loss $\calL_\mathrm{NMSE}$ focuses on the high-energy region, i.e., direct arrival and early reflections, while the EDC loss $\calL_\mathrm{EDC}$ focuses on the reverberation decay.

\subsubsection{Parameterization and Initialization}
All learnable parameters are stored as unconstrained tensors and mapped to valid ranges via nonlinear activations (e.g., sigmoid for the reflection coefficients $\boldsymbol{\alpha}$).
We initialize the reflection coefficients $\boldsymbol{\alpha}$ to $0.5$, source directivity $\Gamma_s$ as omnidirectional, and the global gain $g$ to $1$. The combination coefficients $\boldsymbol{\beta}$ of parametric DART are initialized to have a $0.95\!:\!0.05$ reflection-to-transmission ratio and $0.8\!:\!0.2$ diffuse-to-specular ratio, and the lossless matrix $\bar{\bm{M}}$ of unconstrained DART is initialized to what the parametric model's initial combination coefficients $\boldsymbol{\beta}$ would yield. With these settings, both variants initially predict identical echograms for the same source-receiver pair.

\subsubsection{Optimization Setup}
We use the AdamW optimizer \cite{loshchilov2017decoupled} with a batch size of $1$.
We employ a base learning rate of $10^{-2}$ for the reflection coefficients $\boldsymbol{\alpha}$, source directivity $\Gamma_s$, and global gain $g$, and one fourth of the base learning rate for the lossless matrix $\bar{\bm{M}}$ and combination coefficients $\boldsymbol{\beta}$, so that fine scattering patterns are learned after coarser parameters.
The parameter learning rates are annealed to zero using a cosine scheduler. We optimize DART for $25\si{k}$ steps, which takes no more than $2.5$ hours with a single NVIDIA \texttt{RTX4090} GPU.

\begin{table*}
\setlength\tabcolsep{2.4pt}
\renewcommand{\arraystretch}{.89}
\begin{center}
\caption{Main benchmark results.
Lower is better for all metrics.
For each benchmark setup and metric, the best result is highlighted in bold and blue, and the second-best result is highlighted in pink.
}
\small
\scalebox{0.98}{
\begin{tabular}{llRRRRrRRRRrRRRR}
\toprule
&
& \multicolumn{4}{c}{\bf HAA}
&
& \multicolumn{4}{c}{\bf CR (random)}
&
& \multicolumn{4}{c}{\bf CR (unseen)}
\\
\cmidrule{3-6}
\cmidrule{8-11}
\cmidrule{13-16}

\bf Methods
&
&\multicolumn{1}{r}{\bf {L1}  }
&\multicolumn{1}{r}{\bf {T60} }
&\multicolumn{1}{r}{\bf {EDT} }
&\multicolumn{1}{r}{\bf {C50} }
&
&\multicolumn{1}{r}{\bf {L1}  }
&\multicolumn{1}{r}{\bf {T60} }
&\multicolumn{1}{r}{\bf {EDT} }
&\multicolumn{1}{r}{\bf {C50} }
&
&\multicolumn{1}{r}{\bf {L1}  }
&\multicolumn{1}{r}{\bf {T60} }
&\multicolumn{1}{r}{\bf {EDT} }
&\multicolumn{1}{r}{\bf {C50} }
\\

\midrule
Nearest neighbor & &
$0.895$ & $7.53$ & $0.220$ & $1.60$ & & $0.891$ & \cellcolor{2ndcol!30} $5.01$ & $0.179$ & $1.50$ & &  $-$ & $-$ & $-$ & $-$
\\
Linear interpolation & &
$0.892$ & \cellcolor{2ndcol!30} $7.32$ & $0.206$ & \cellcolor{2ndcol!30} $1.57$ & & $0.983$ & \cellcolor{1stcol!30} $\bf 4.85$ & $0.184$ & $1.57$ & &  $-$ & $-$ & $-$ & $-$
\\
\midrule
DiffRIR \cite{wang2024hearing} & &
$0.745$ & $8.13$ & $0.160$ & $1.65$ & & $0.772$ & $14.06$ & $0.217$ & $1.86$ & & $5.930$ & $42.88$ & $0.349$ & $4.46$
\\
\midrule
INRAS \cite{su2022inras} & &
$0.989$ & $10.75$ & $0.331$ & $2.78$ & & $1.838$ & $18.34$ & $0.328$ & $3.03$ & & $6.726$ & $61.22$ & $1.934$ & $4.75$
\\
\quad w/ echogram & &
$0.822$ & $8.60$ & $0.232$ & $1.81$ & & $2.574$ & $15.34$ & $0.428$ & $3.17$ & & $1.757$ & $22.17$ & $0.542$ & $2.93$
\\
\midrule
AVR \cite{lan2024acoustic} & &
$0.972$ & $9.18$ & $0.287$ & $2.09$ & & $1.049$ & $9.32$ & $0.329$ & $2.53$ & & $2.985$ & $25.46$ & $1.030$ & $9.84$
\\
\quad w/ echogram & &
$0.948$ & $8.48$ & $0.259$ & $2.18$ & & $1.236$ & $8.37$ & $0.277$ & $2.86$ & & $3.509$ & \cellcolor{1stcol!30} $\bf 15.35$ & $0.647$ & $5.03$
\\
\midrule
DART (UC) & &
\cellcolor{2ndcol!30} $0.642$ & \cellcolor{1stcol!30} $\bf 7.00$ & \cellcolor{2ndcol!30} $0.115$ & \cellcolor{1stcol!30} $\bf 1.12$ & & \cellcolor{1stcol!30} $\bf 0.510$ & $5.07$ & \cellcolor{1stcol!30} $\bf 0.080$ & \cellcolor{1stcol!30} $\bf 0.72$ & & \cellcolor{1stcol!30} $\bf 0.831$ & \cellcolor{2ndcol!30} $19.90$ & \cellcolor{1stcol!30} $\bf 0.073$ & \cellcolor{1stcol!30} $\bf 0.69$
\\
DART (P) & &
\cellcolor{1stcol!30} $\bf 0.638$ & $15.95$ & \cellcolor{1stcol!30} $\bf 0.097$ & $1.84$ & & \cellcolor{2ndcol!30} $0.577$ & $5.82$ & \cellcolor{2ndcol!30} $0.083$ & \cellcolor{2ndcol!30} $0.77$ & & \cellcolor{2ndcol!30} $0.900$ & $24.51$ & \cellcolor{2ndcol!30} $0.125$ & \cellcolor{2ndcol!30} $1.82$
\\

\arrayrulecolor{black}
\bottomrule
\end{tabular}
}
\label{table:results-main}
\end{center}
\end{table*}

\subsection{Baselines}
\subsubsection{Low Anchors}
Following prior works \cite{su2022inras, lan2024acoustic, wang2024hearing}, we use nearest neighbor and linear interpolation as the simplest baselines.

\subsubsection{DiffRIR}
DiffRIR is a differentiable image-source method (ISM) introduced with the HAA dataset \cite{wang2024hearing}.
It models each RIR as an early specular component and a late reverberation residual. The early component is computed via the ISM, where the energy of each image-source path is determined by surface reflection coefficients, source directivity, and air absorption. The late reverberation is a learnable residual signal shared across the scene.
We adopt the official code with minor hyperparameter adjustments, e.g., sampling rate, to match our benchmark setup.

\subsubsection{INRAS}
Inspired by ART, INRAS \cite{su2022inras} employs three neural network encoders corresponding to injection, reflections, and detection.
These encoders take ``bounce points'' sampled from the room geometry, or their relative positions with respect to the source and receiver, as input.
Then, a neural decoder aggregates the latent features from the encoders and predicts the RIR.
Starting from the official implementation, we adapt it to our benchmark setup: bounce points are taken from the (subdivided) mesh patch centers, source orientation is provided to the injection encoder, and the stereo ``listener module'' is removed for mono output.
We train INRAS with a batch size of $4$, a learning rate of $5\times 10^{-5}$, and $125\si{k}$ optimization steps.

\subsubsection{AVR}
AVR extends differentiable volume rendering in vision \cite{mildenhall2021nerf} to acoustics by accounting for the nonnegligible speed of sound.
An RIR is modeled as the aggregation of contributions from all incoming directions at the receiver.
For each direction, a ray is cast into the scene and contributions are accumulated along the ray based on a neural-network-predicted volumetric density and source signal, accounting for propagation delay, distance-based attenuation, and transmittance.
Unlike the other baselines, AVR does not take explicit room geometry as input.
Since AVR tends to overfit on sparse measurements, we train it for $2.5\si{k}$ steps with a $2\times 10^{-4}$ learning rate.

\subsubsection{Echogram Variants}
The neural baselines, INRAS and AVR, were originally proposed for RIR prediction, which could be more challenging than echogram prediction. Therefore, for a fair comparison, we also train echogram variants, modified to predict energy quantities, e.g., using softplus activation for non-negativity, and trained with the same echogram loss $\mathcal{L}_\mathrm{Echogram}$ as DART. We denote these variants with ``w/ echogram'' or ``(E).''
We do not include an echogram variant of DiffRIR, as it did not yield improved results in preliminary experiments. We keep the same hyperparameters for the echogram variants, except for AVR (E), which is trained for $50\si{k}$ steps, longer than the RIR-predicting AVR, as it exhibits less overfitting.

\begin{figure*}[t]
\centering
\hspace{-15.1mm}
\includegraphics[clip, trim={0mm -.4mm 0mm 0mm}, width=.912\linewidth]{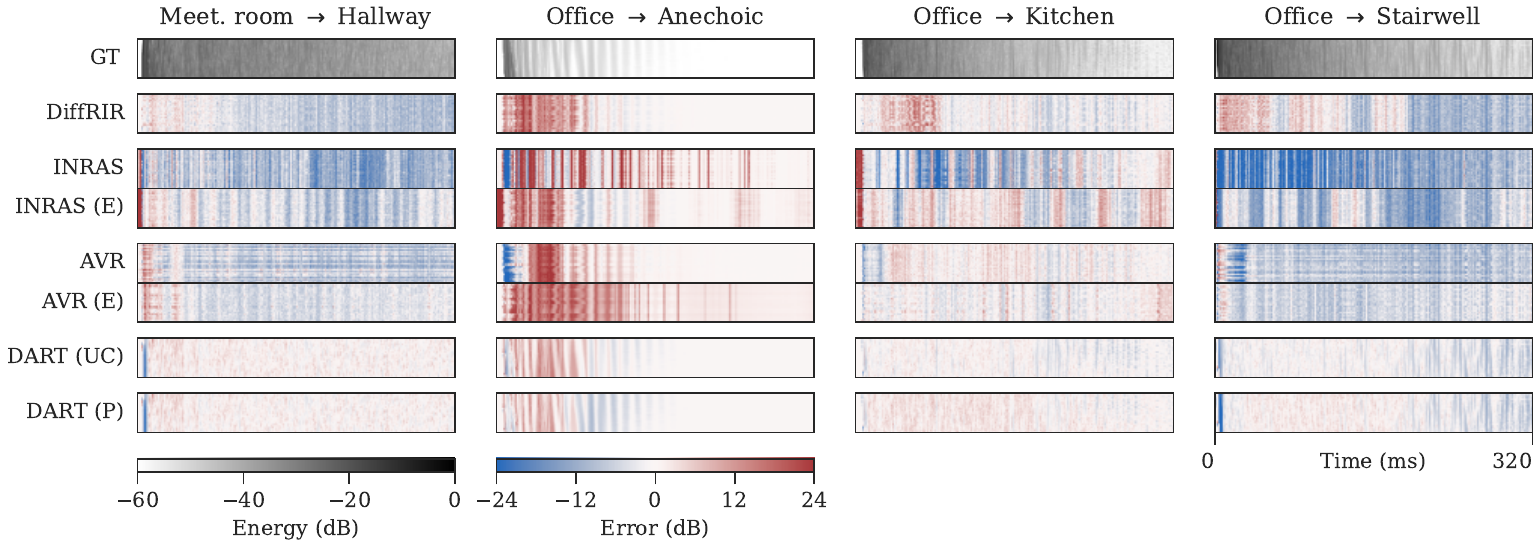}
\\
\setlength\tabcolsep{5.3pt}
\centering
\hspace{11.9mm}
\begin{tabular}{ccccc}
{
\includegraphics[clip, trim={2.9cm 1.7cm .7cm 2.5cm}, width=.181\linewidth]{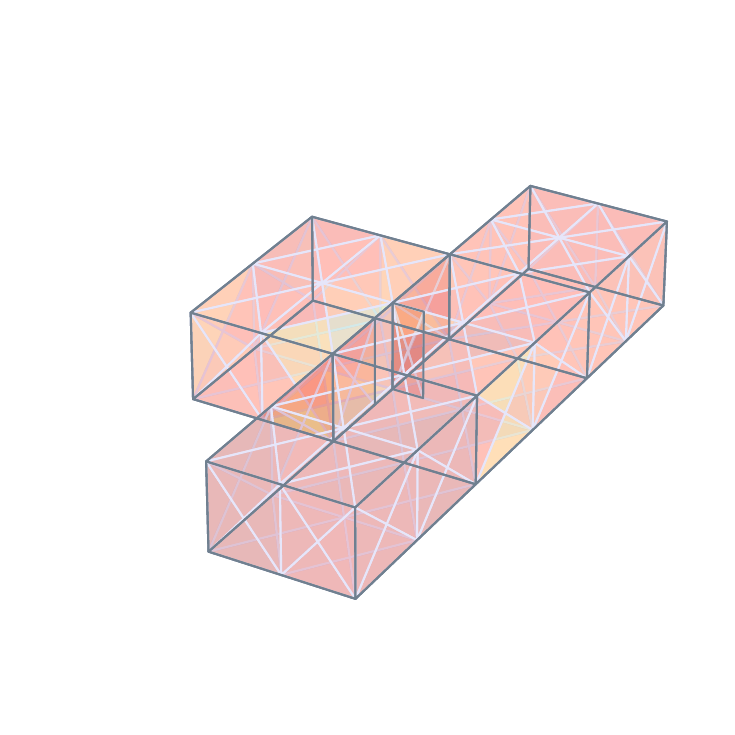}
}
&
{
\includegraphics[clip, trim={1cm 0.95cm .2cm .6cm}, width=.181\linewidth]{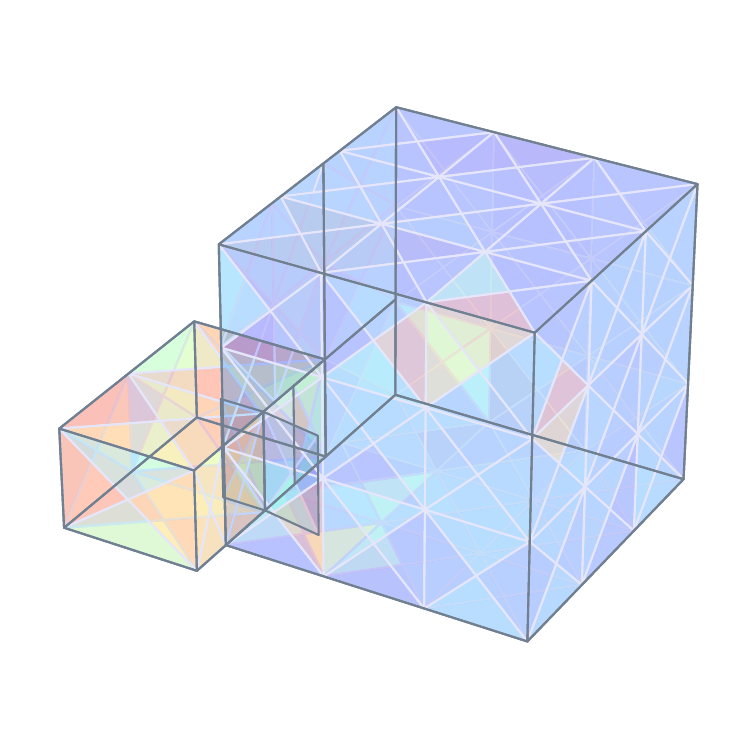}
}
{
\includegraphics[clip, trim={1.8cm 1.7cm .8cm 2cm}, width=.185\linewidth]{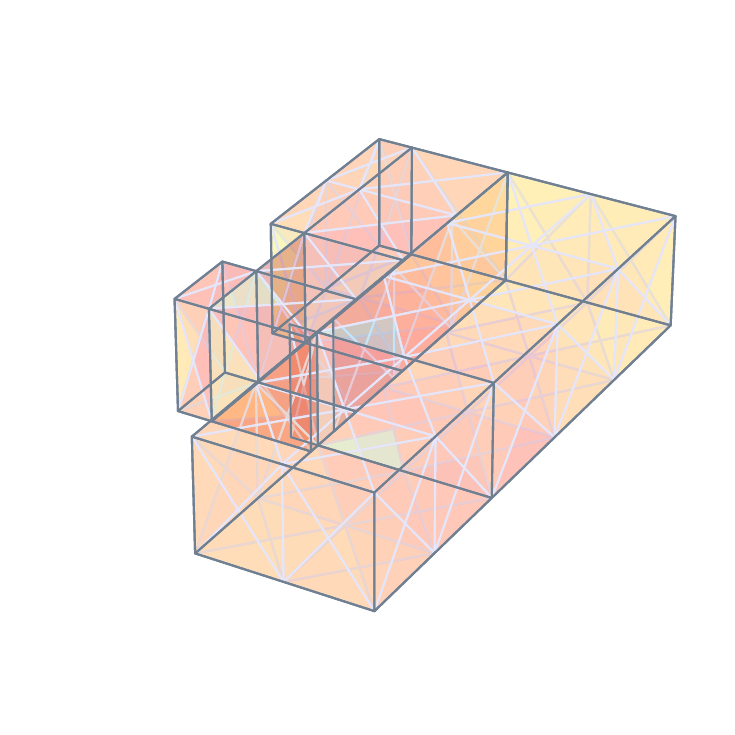}
}
&
{
\includegraphics[clip, trim={1.55cm .3cm 0.25cm 2.1cm}, width=.181\linewidth]{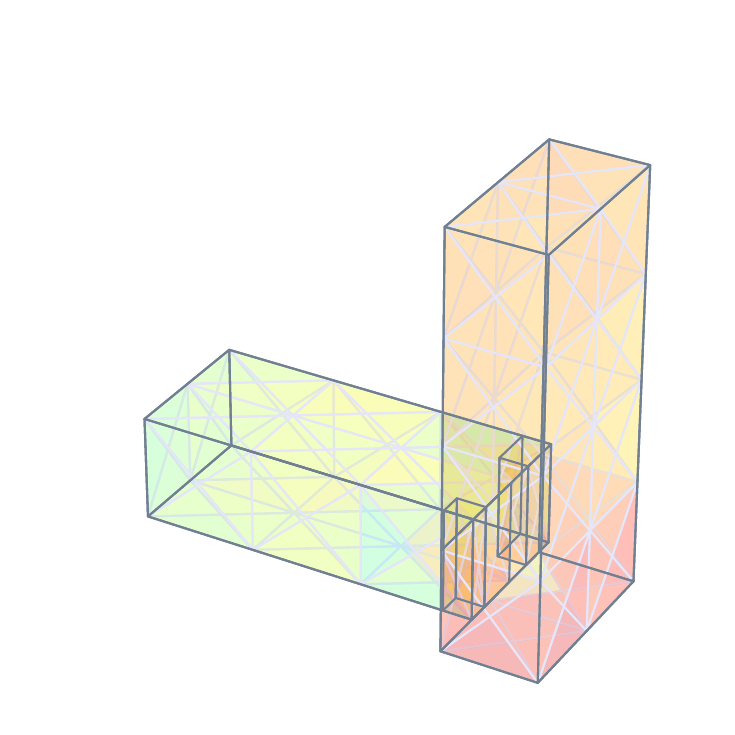}
}
&
\includegraphics[clip, trim={3.9cm 0.1cm .3cm 0.25cm}, width=.056\linewidth]{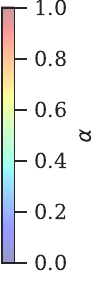}
\end{tabular}
\caption{
Evaluation results on the Coupled Room (CR) dataset, unseen split.
The top row shows all ground-truth test echograms, in decibels $\smash{10\log_{10} \hat{E}^\mathrm{true}[n]}$, stacked vertically.
For each model, the signed residuals of its predictions, $\mathcal{E} = 10\log_{10}\hat{E}^\mathrm{pred}[n] - 10\log_{10}\hat{E}^\mathrm{true}[n]$, are shown. 
All residual plots share the same diverging colormap with fixed limits of $\pm 24 \si{dB}$, centered at zero.
White corresponds to a perfect match (zero error), while red and blue indicate over- and underestimation of the echogram, respectively.
We also visualize the optimized reflection coefficients $\smash{\alpha_1, \cdots, \alpha_{\Npat}}$ of parametric DART (bottom row), with higher reflection coefficients (more reverberant) shown in red and lower reflection coefficients (more absorptive; anechoic) shown in blue.
}
\label{fig:qualitative-unseen}
\end{figure*}

\section{Results}
\subsection{Main Results}
Table \ref{table:results-main} summarizes the average evaluation results for each benchmark.
Per-scene results are reported in the supplementary material.

Both DART variants perform well across all three benchmarks, with a wider margin over the baselines on the more challenging CR benchmarks. The unconstrained (UC) variant generally achieves the strongest results, while the parametric (P) variant remains competitive despite its constrained material model. For reverberation time, the results are more mixed: the unconstrained variant does not always lead but remains close, while the parametric variant occasionally struggles. DART's advantage is largest in the remaining metrics, which reflect acoustic characteristics that vary considerably with source and receiver position and orientation.

The neural baselines, INRAS and AVR, struggle to surpass even simple nearest-neighbor and linear interpolation.
While this contrasts with prior results on larger-scale benchmarks \cite{chen2020soundspaces, chen2024real} where orders of magnitude more data were available, it is consistent with more recent findings under sparse conditions \cite{wang2024hearing, liu2025hearing}, suggesting that these models lack sufficient inductive bias to produce physically plausible predictions from limited data.
The echogram variants generally show slightly better results than their RIR counterparts on the HAA and CR (unseen) benchmarks, while the comparison is mixed on CR (random).

\subsection{Qualitative Analysis}
The CR (unseen) setup is the most challenging among the three benchmarks, as it requires models to generalize to entirely new source and receiver regions. We examine its test results in detail.
In Figure \ref{fig:qualitative-unseen}, for each scene, the test echograms (in decibels) are vertically stacked and visualized as a matrix, where each echogram corresponds to a row.
Then, signed residuals of the model predictions are shown, each defined as
\begin{equation}
    \mathcal{E} = \smash{10\log_{10} \hat{E}^\mathrm{pred}[n] - 10\log_{10} \hat{E}^\mathrm{true}[n]}.
\end{equation}

It reveals that the neural baselines, regardless of the optimization target (RIR or echogram), fail to accurately reconstruct both early and late parts of the echogram.
For example, INRAS, despite its architecture being inspired by ART and having access to geometry and source/receiver positions,
was unable to estimate direct arrivals, possibly because the corresponding delay lengths were unseen during training.
Among the baselines, DiffRIR performs best, owing to the physical grounding of its ISM backbone. However, its early reflection model captures only specular paths, missing scattered energy, while its single shared residual for late reverberation cannot adapt to the varying reverberation decays across different source-receiver pairs.
In contrast, DART, regardless of the parameterization method, consistently achieves lower errors than any baseline.

We observe that the baseline methods especially struggle at the \texttt{Office\;$\to$\;Anechoic} and \texttt{Office\;$\to$\;Stairwell} scenes.
These scenes comprise two coupled rooms with drastically different acoustic properties, e.g., \texttt{Anechoic} having much lower reverberation time than \texttt{Office}.
The baselines, trained on measurements from receivers only in \texttt{Office}, did not capture this and overestimated the decay.
DART, by contrast, recovered markedly different reflection coefficients for the two rooms and, as a result, retained its accuracy for those challenging scenes.
For the other two scenes
where all the baselines performed relatively well, DART predicted nearly homogeneous reflection coefficients across the two rooms.
Notably, DART recovered this difference without any receiver inside \texttt{RoomB}, e.g., \texttt{Anechoic} in \texttt{Office\;$\to$\;Anechoic}, during optimization.

\begin{figure*}
    \centering
    \includegraphics[clip, trim={0mm 3mm 0mm 0.1mm}, width=.997\linewidth]{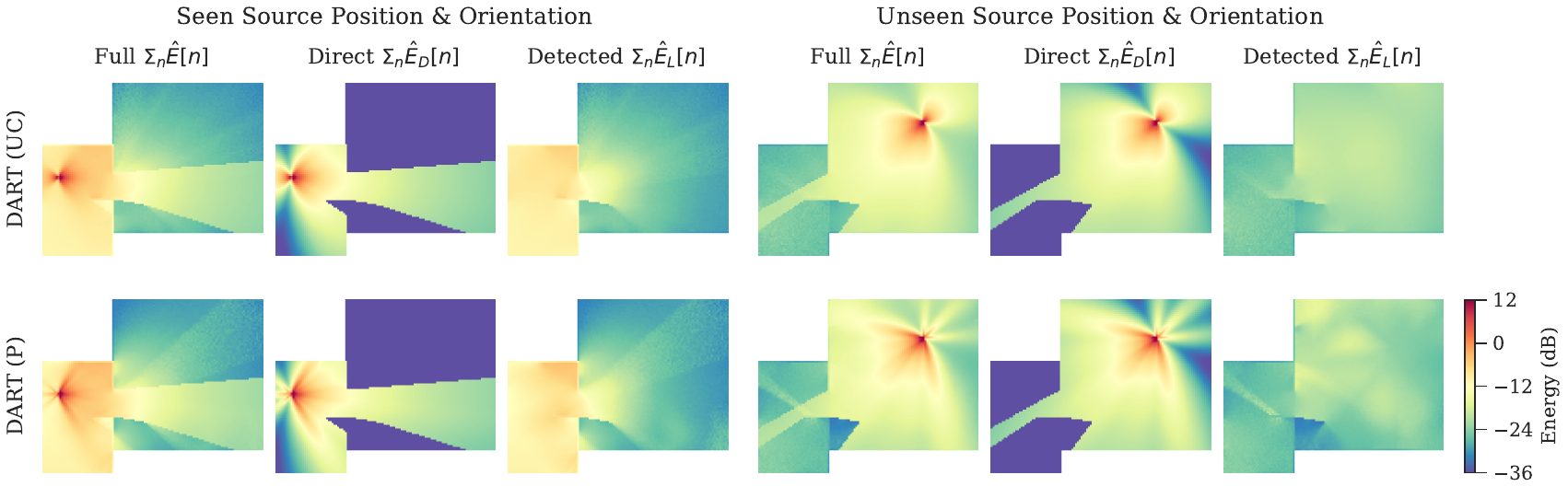}

    \caption{
    Spatial energy distribution predicted by the unconstrained and parametric DART for the \texttt{Office$\;\to\;$Anechoic} scene.
    Results are shown for two different source configurations. The first (left three columns) corresponds to a source located in \texttt{Office} and oriented to the right, which is observed during optimization. The second (right three columns) corresponds to an unseen source located in \texttt{Anechoic} and oriented toward the lower left.
    We also visualize the direct-arrival and detected energy distributions separately; their sum yields the full energy distribution.
    }
    \label{fig:energy-distribution}
\end{figure*}

\subsection{Material Parameterization Comparison}
The evaluation results indicate that the unconstrained DART generally outperforms the parametric DART.
To further investigate the effects of different material parameterizations, we visualize the spatial energy distributions predicted by these two variants.
We choose the \texttt{Office$\;\to\;$Anechoic} scene from the CR dataset (random split), and the full energy is computed as a sum of the predicted echogram $\smash{\hat{E}[n]}$ over time $n$. We also visualize the sum of direct arrival $\smash{\hat{E}_D[n]}$ and detected energy components $\smash{\hat{E}_L[n]}$ separately. Two source configurations are evaluated: one seen source in \texttt{Office} and one unseen source in \texttt{Anechoic}. Receivers are positioned on a regular grid with $0.1\si{m}$ spacing at $1.5\si{m}$ height.

Figure \ref{fig:energy-distribution} shows the results. Regardless of the parameterization, both DART variants respect the surrounding room geometry and predict valid energy distribution fields. However, we observe several differences. First, the source directivity of the parametric DART deviates slightly from typical radiation patterns. Second, in the unseen configuration, parametric DART exhibits irregular patterns in the scattered field produced by the radiances. We suspect that this is caused by the model trying to match the small number of measurements in the local receiver region (the small transition area between \texttt{Office} and \texttt{Anechoic}) with limited degrees of freedom.
In contrast, the unconstrained DART produces smoother source radiation and energy fields. It also has fewer visible discontinuities caused by the patch and direction discretizations.

\begin{figure}[t]
\centering
\includegraphics[clip, trim={1.2mm 6mm 5mm 0mm}, width=\linewidth]{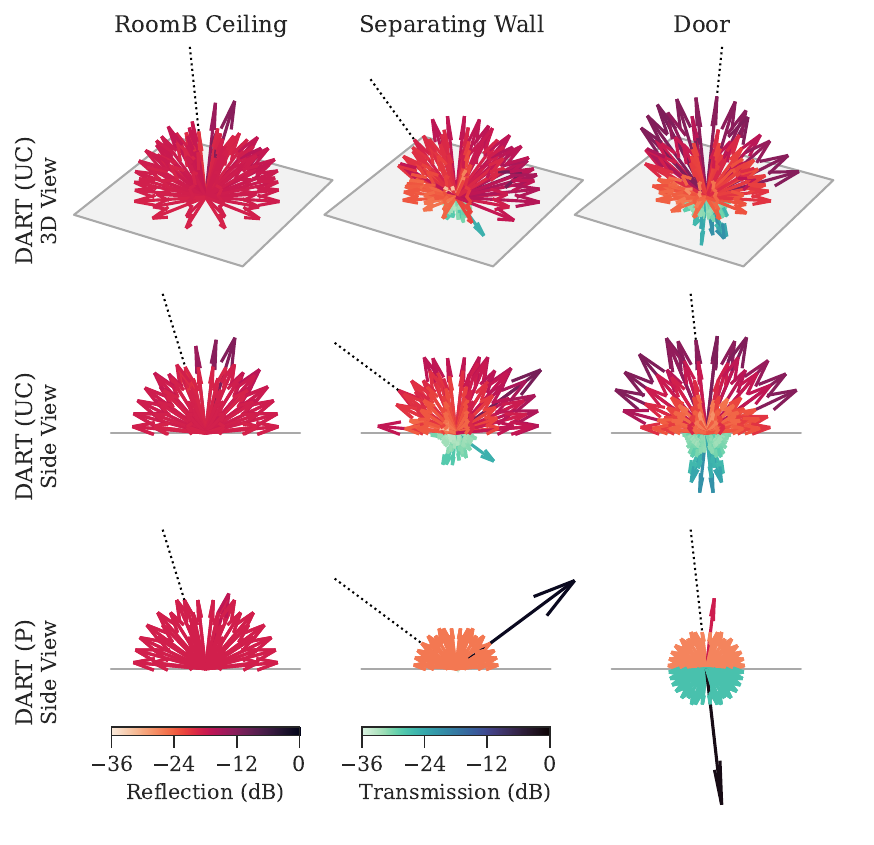}

\caption{
Example reflection and transmission patterns of the lossless material matrices of optimized unconstrained and parametric DART. Three representative surface patches from the $\texttt{MeetingRoom}\!\to\!\texttt{Hallway}$ scene are shown, each corresponding to one column. Each black dotted line indicates the incident direction. The colors and lengths of the arrows in the reflected and transmitted patterns are shown on a decibel scale. Surface patches are visualized as rectangles for clarity, although they are triangular and vary in shape and size in the actual geometry.
The materials of the unconstrained DART are shown in a three-dimensional view (top row) and a side view (middle row), while the materials of the parametric DART are shown only in a side view (bottom row) for brevity.
}
\label{fig:example-materials}
\end{figure}

We also visualize the learned scattering patterns of representative surface patches of the \texttt{MeetingRoom$\;\to\;$Hallway} scene.
Because each lossless material matrix encodes scattering responses for all incident directions, we sample a single row, corresponding to a single incident direction, from the matrix for each patch. Figure~\ref{fig:example-materials} shows the results.
The unconstrained DART converges to a wide range of scattering patterns, such as broadened or anisotropic lobes, whereas the parametric DART is limited to combinations of predefined BSDFs. It produces noticeably different, possibly less realistic, reflection-to-transmission ratios in some patches, likely a consequence of fitting the observations under its constrained parameterization.

\section{Further Analysis}
We conduct additional experiments to further analyze DART.
All eight scenes from the HAA and CR datasets (random split) were used for evaluation, and we report the average results.
We retain all baselines except the RIR-predicting versions of the neural models, as they did not improve over their echogram counterparts in the main benchmarks.

\subsection{Effect of Measurement Quantity}
\looseness=-1
We assess the effect of measurement quantity by optimizing the models with extremely sparse to moderate numbers of measurements, ranging from $4$ to $48$.

\begin{figure*}[t]
\centering
    \hspace{-7mm}
    \includegraphics[clip, trim={0cm 0.16cm 0cm 0.01cm}, width=.83\linewidth]{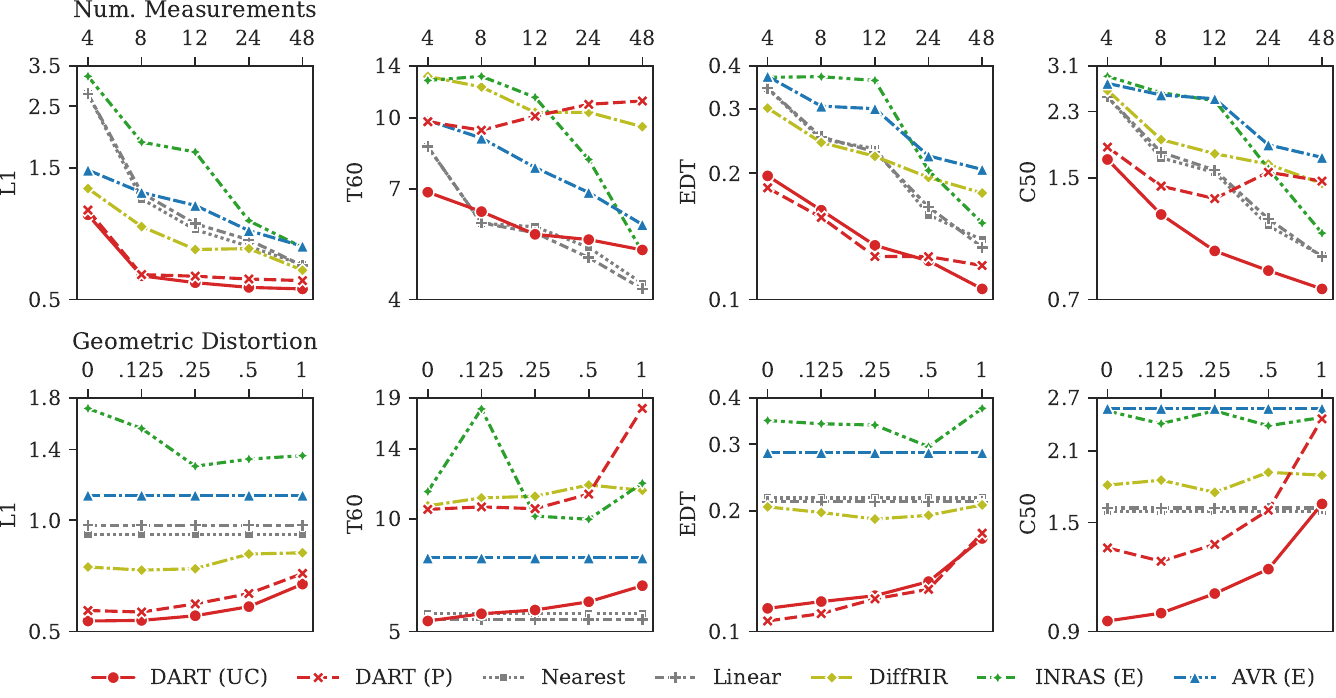}
    \caption{Test results with different numbers of measurements (top row) and geometric distortion (bottom row).
    }
    \label{fig:main-data-and-distortion}
\end{figure*}

The top row of Figure \ref{fig:main-data-and-distortion} shows the results.
Most models exhibit an expected trend: more data yields improvement across all metrics.
The unconstrained DART shows consistently strong performance across the entire range.
The parametric DART, by contrast, shows mixed results: while competitive in some metrics under data-scarce conditions, it underperforms in others, e.g., reverberation time and clarity, with this gap widening as more measurements become available.

\subsection{Sensitivity to Geometry Distortion}
DART depends on the room geometry, which in practice is known only approximately. The benchmarks above already use coarse meshes that omit fine architectural detail, yet DART, especially the unconstrained variant, achieved strong results.
To further stress-test this robustness, following prior works \cite{wang2024hearing, gao2024soaf}, we optimized DART and other baselines on synthetically distorted meshes, created by perturbing each vertex by uniform noise $u \sim \mathcal{U}[-s, s)$, varying the scale $s$ from $0.125\si{m}$ to $1\si{m}$. The echogram measurements are kept unchanged.

The bottom row of Figure \ref{fig:main-data-and-distortion} shows the results.
The unconstrained DART retains its performance well up to $0.25\si{m}$ distortion, likely because the low radiance sampling rate ($1\si{kHz}$ $\approx 0.34\si{m}$ spatial resolution) makes sub-resolution perturbations negligible and the flexible material matrix can adapt to modest geometric changes.
At the largest distortion ($s=1\si{m}$), however, the gap to even the simplest baselines shrinks, as such perturbations effectively erase critical geometric features (e.g., room dimensions, aperture sizes).
The parametric DART is more sensitive to severe distortion due to its limited flexibility.
DiffRIR exhibits a similar trend but with less degradation, potentially due to its learnable residual signal.
INRAS (E), despite taking geometry as input, does not show a clear trend, suggesting that its ART-inspired architecture might not utilize the geometry in the same way as ART itself.
The remaining baselines do not rely on geometry and are therefore unaffected.

\subsection{Ablation Study}
We evaluate the various design choices and hyperparameters of DART; Table \ref{table:ablation} reports the results, along with the parameter count, memory usage (in GB), and iteration time (in ms) for each variant during training and inference (measured with a single \texttt{RTX4090} GPU). In addition to the base $12$-measurement setup, we include a $48$-measurement setup to investigate the effects of data availability on each configuration.

First, we verify that allowing each subpatch to have its own independent material parameters is advantageous, as the shared variant gave degraded results in most metrics, especially in T60 and C50.
This gap was particularly pronounced for the parametric DART, which has orders of magnitude fewer parameters than the unconstrained one, so further restriction was especially detrimental.

Next, we varied the discretization resolution for the unconstrained variant.
For the patch resolution, we roughly halved or doubled the number of patches by adjusting the mesh subdivision parameters.
For the directional resolution, we explored $8\times 8$ and $16\times 16$ grids, compared to the base $12\times 12$.
With $12$ measurements, resolution has no clear effect on performance.
However, with $48$ measurements, finer patch and directional resolutions consistently improve the results, suggesting that the optimal discretization depends on data availability.
That said, increasing the resolution also comes with increased computational cost, and the gains from finer discretization may not always justify this cost; the most lightweight variant, with both coarser patch and directional resolutions, still achieves competitive results with significantly fewer parameters and faster runtime.
The base setting offers a good balance between computational cost and performance.

\begin{table*}
\caption{
    Ablation study of DART, along with the parameter count, memory usage (in GB), and wall-clock time (in ms). 
    Costs for both training and inference are shown, with the latter in parentheses.
}
\label{table:ablation}
\centering
\scalebox{0.97}{
\small
\setlength\tabcolsep{2.4pt}
\renewcommand{\arraystretch}{.89}
\begin{tabular}{llX{17mm}X{17.7mm}X{20.2mm}lX{9.2mm}X{9.2mm}X{9.2mm}X{9.2mm}lX{9.2mm}X{9.2mm}X{9.2mm}X{9.2mm}}
\toprule
&
& \multicolumn{3}{c}{\bf Model Size \& Runtime}
&
& \multicolumn{4}{c}{\bf 12 Measurements}
&
& \multicolumn{4}{c}{\bf 48 Measurements}
\\
\cmidrule{3-5}
\cmidrule{7-10}
\cmidrule{12-15}
\bf Variants
& &
\multicolumn{1}{r}{\bf {Parameters}  }
& \multicolumn{1}{r}{\bf {Memory}  }
& \multicolumn{1}{r}{\bf {Speed}}
& &
\multicolumn{1}{r}{\bf {L1}  }
&\multicolumn{1}{r}{\bf {T60} }
&\multicolumn{1}{r}{\bf {EDT} }
&\multicolumn{1}{r}{\bf {C50} }
& &
\multicolumn{1}{r}{\bf {L1}  }
&\multicolumn{1}{r}{\bf {T60} }
&\multicolumn{1}{r}{\bf {EDT} }
&\multicolumn{1}{r}{\bf {C50} }
\\
\midrule
\arrayrulecolor{lightgray}

\rowcolor{gray!20}
DART (UC) & &
$4204555$ & $3.32$ $(0.74)$ & $143.4$ $(56.6)$
& & $0.576$ & $6.04$ & $0.097$ & $0.92$
& & $0.547$ & $5.58$ & $0.071$ & $0.72$
\\
\midrule
\quad Shared parameters & &
$4204369$ & $3.35$ $(0.74)$ & $143.3$ $(56.1)$
& & $0.572$ & $9.84$ & $0.091$ & $1.23$
& & $0.550$ & $8.61$ & $0.073$ & $0.96$
\\
\midrule
\quad Coarser patch & &
$2545595$ & $2.34$ $(0.48)$ & $91.3$ $(35.9)$
& & $0.576$ & $6.07$ & $0.101$ & $0.94$
& & $0.551$ & $5.92$ & $0.066$ & $0.77$
\\
\quad Coarser direction & &
$830795$ & $1.66$ $(0.29)$ & $44.1$ $(17.8)$
& & $0.579$ & $6.44$ & $0.094$ & $0.92$
& & $0.558$ & $5.91$ & $0.070$ & $0.80$
\\
\quad Coarser pat. \& dir. & &
$503035$ & $1.26$ $(0.19)$ & $34.2$ $(13.4)$
& & $0.579$ & $6.55$ & $0.097$ & $0.95$
& & $0.561$ & $6.05$ & $0.069$ & $0.84$
\\
\midrule
\quad Finer patch & &
$7885373$ & $5.74$ $(1.38)$ & $293.3$ $(121.6)$
& & $0.571$ & $7.22$ & $0.101$ & $0.98$
& & $0.541$ & $5.53$ & $0.071$ & $0.70$
\\
\quad Finer direction & &
$13287755$ & $5.72$ $(1.56)$ & $409.3$ $(160.6)$
& & $0.577$ & $6.04$ & $0.098$ & $0.91$
& & $0.543$ & $5.49$ & $0.070$ & $0.71$
\\
\arrayrulecolor{black}
\midrule
\arrayrulecolor{lightgray}
\rowcolor{gray!20}
DART (P) & &
$971$ & $3.17$ $(0.67)$ & $141.1$ $(55.8)$
& & $0.608$ & $10.89$ & $0.090$ & $1.30$
& & $0.586$ & $11.75$ & $0.084$ & $1.46$
\\
\midrule
\quad Shared parameters & &
$198$ & $3.17$ $(0.67)$ & $141.3$ $(55.7)$
& & $0.624$ & $15.23$ & $0.110$ & $2.44$
& & $0.583$ & $13.35$ & $0.100$ & $1.47$
\\
\midrule
\quad Specular refl. only & &
$534$ & $3.14$ $(0.63)$ & $141.1$ $(55.8)$
& & $0.678$ & $10.81$ & $0.113$ & $1.66$
& & $0.636$ & $9.85$ & $0.106$ & $1.41$
\\
\quad Diffusive refl. only & &
$534$ & $3.14$ $(0.63)$ & $140.7$ $(55.9)$
& & $0.621$ & $13.24$ & $0.107$ & $1.62$
& & $0.601$ & $12.47$ & $0.096$ & $1.51$
\\
\midrule
\quad w/o mat. decomp. & &
$971$ & $2.96$ $(1.13)$ & $724.8$ $(209.0)$
& & $0.604$ & $11.69$ & $0.091$ & $1.24$
& & $0.589$ & $12.01$ & $0.086$ & $1.44$
\\
\quad Explicit injection & &
$971$ & $3.17$ $(0.67)$ & $140.9$ $(56.9)$
& & $0.612$ & $11.50$ & $0.091$ & $1.37$
& & $0.591$ & $11.42$ & $0.091$ & $1.43$
\\
\arrayrulecolor{black}
\bottomrule

\end{tabular}
}
\end{table*}

We also examined special cases of the parametric DART: optimizing the reflection coefficients under fixed scattering properties (only specular or only diffuse reflection), as in prior works \cite{liu2025hearing, zhi2023differentiable, tang2020scene, li2018scene, schissler2017acoustic}.
Both restrictions degraded the results, confirming the importance of flexible material properties.

\looseness=-1
Finally, we validated the two nontrivial approximations that we introduced for DART: the visibility-material decomposition \eqref{eq:scattering-decomposition} and the use of the material matrix for injection \eqref{eq:injection-wo-monte-carlo}.
For the parametric variant, whose predefined component BSDFs are known, each can be disabled: the decomposition by precomputing the full matrix \eqref{eq:scattering-kernel} per component, and the material-matrix injection by reverting to explicit single-bounce ray tracing.
For the unconstrained variant, disabling either would require explicit BSDF parameterization and numerical integration, significantly increasing computational cost and system complexity.
We therefore test on the parametric variant only.

The last two rows of Table \ref{table:ablation} report the results.
Removing the decomposition shows no clear performance difference but incurs a significant slowdown, as the full matrix is much denser than its decomposed factors.
Using explicit injection slightly degrades most metrics, likely due to increased stochasticity from ray tracing.
Overall, these results confirm that both approximations are reasonable, while the decomposition additionally provides a clear computational advantage.

\subsection{Computational Cost}\label{sec:cost}
\begin{table}
\caption{Asymptotic work complexity and execution speed (in ms) of each operation in DART.}
\centering
\label{table:speed}
\setlength\tabcolsep{2.4pt}
\renewcommand{\arraystretch}{.96}
\small
\scalebox{0.97}{
\begin{tabular}{llcrr}
\toprule
\bf Operation
& &
\bf Work Complexity & &
\bf Speed
\\
\midrule
Precomputation
& & $\smash{\approx  \mathcal{O}(N_\mathrm{precompute} N_\mathrm{pat} \log N_\mathrm{pat})}$ & & $\bf 406.8$ \\
\midrule
\arrayrulecolor{lightgray}
Injection & & $\smash{\approx \mathcal{O}(N_\mathrm{ray} \log N_\mathrm{pat} + \bar{N}_\mathrm{mat} T) }$ & & $5.2$ \\
\midrule
Delay
& & $\smash{\mathcal{O}(N_\mathrm{order} \bar{N}_\mathrm{rad} T) }$ & & $2.6$ \\
Visibility
& & $\smash{\mathcal{O}(N_\mathrm{order} \bar{N}_\mathrm{vis} T) }$ & & $12.4$ \\
Material
& & $\smash{\mathcal{O}(N_\mathrm{order} \bar{N}_\mathrm{mat} T) }$ & & $33.4$ \\
Accumulation
& & $\smash{\mathcal{O}(N_\mathrm{order} \bar{N}_\mathrm{rad} T) }$ & & $2.6$ \\
\midrule
FFT \& IFFT & & $\smash{\mathcal{O}(\bar{N}_\mathrm{rad} T \log T )}$ & & $0.2$
\\
\midrule
Detection & & $\smash{\approx \mathcal{O}(N_\mathrm{ray} \log N_\mathrm{pat} + \bar{N}_\mathrm{rad} T \log T)}$ & & $5.8$ \\
\arrayrulecolor{black}
\midrule
Total (sync) & & $-$ & & $62.2$\\
\arrayrulecolor{lightgray}\midrule\arrayrulecolor{black}
Total (wall clock) & & $-$ & & $\bf 56.6$\\
\bottomrule
\end{tabular}
}
\end{table}

Table \ref{table:speed} lists the asymptotic work complexity and measured execution time of each operation in the unconstrained DART.
Operations marked with the approximation symbol $\approx$ involve ray tracing, whose exact complexity depends on the scene geometry and the implementation; the $\log N_\mathrm{pat}$ factor reflects acceleration-structure traversal for ray-triangle intersection.

\looseness=-1
The precomputation of fixed components such as the mean visibility matrix and propagation delays is performed once per scene before optimization.
The visibility-material decomposition \eqref{eq:scattering-decomposition} reduces the four nested integrals (or three, if we collapse the Dirac delta) of the full scalar-valued matrix \eqref{eq:scattering-kernel} to at most two surface integrals for the mean visibility matrix \eqref{eq:mean-visibility-matrix}. Precomputing these decomposed factors is therefore substantially cheaper than precomputing the full matrix, but would still dominate the overall training time unless performed beforehand.

The main iterative radiance calculation is the most costly part.
Each of the $N_\mathrm{order}$ reflection steps is sequential, as it depends on the previous step's output. Within each step, the batched sparse matrix multiplications are performed in parallel.
The sparsity ratios, averaged across all tested scenes, are
\begin{subequations}\label{eq:sparsity}
\begin{align}
    \bar{N}_\mathrm{rad} &\approx 5.8\times 10^{-1} N_\mathrm{rad}, \\
    \bar{N}_\mathrm{vis} &\approx 8.9\times 10^{-4}\bar{N}_\mathrm{rad}^2 = 3.0\times 10^{-4} N_\mathrm{rad}^2 , \\
    \bar{N}_\mathrm{mat} &\approx 5.6\times 10^{-3}\bar{N}_\mathrm{rad}^2 = 1.9\times 10^{-3} N_\mathrm{rad}^2.
\end{align}
\end{subequations}
Applying the material matrix is the most costly step, as this matrix is roughly six times denser than the mean visibility matrix. The delay and accumulation steps are relatively lightweight, as they involve only elementwise multiplications and additions on the frequency-domain radiances, respectively.

Injection and detection each involve one set of ray-triangle intersection tests, parallelized over $N_\mathrm{ray}$ rays. The detection and direct-arrival tests are fused into a single pass.
Finally, the $T\log T$ terms correspond to the FFT and IFFT required for the frequency-domain radiance computation and the frequency-domain convolution for detection, respectively.

The per-operation times reported in Table~\ref{table:speed} are measured with GPU synchronization barriers. The actual wall-clock time is $56.6\si{ms}$, slightly shorter than the sum of the per-operation times, as the GPU pipelines consecutive operations.

Inference time is scene-dependent, mainly driven by the truncation order and number of radiances: $87\si{ms}$ for \texttt{Complex}, $34\si{ms}$ for \texttt{Dampened}.
Overall, the inference time of DART is higher than that of the other baselines, approximately $5\si{ms}$ except for AVR at $11.5\si{ms}$.
Still, this gap can be narrowed by reducing the discretization resolution, which drops inference time to $13.4\si{ms}$ with only a modest performance trade-off (see Table~\ref{table:ablation}).
We can further apply ART acceleration techniques after optimization, e.g., precomputing and compressing the transport matrix \cite{cao2016interactive, scerbo2024mod}.
When evaluating multiple receivers for the same source, DART can reuse the computed radiances and perform only separate detections, reducing per-query cost.

\section{Conclusion}
This paper introduced differentiable acoustic radiance transfer (DART).
By decomposing the reflection kernel into mean visibility and material matrices, computing radiance via a frequency-domain truncated Neumann series, and exploiting sparsity, we achieved tractable end-to-end optimization on a single GPU.
Experiments on real measurements showed that DART performs favorably against existing baselines, especially in various challenging yet practical scenarios.

We presented two variants: unconstrained DART, which learns free-form material matrices, and parametric DART, which constrains them to convex combinations of predefined BSDFs.
The unconstrained variant generally achieved stronger results than the parametric variant owing to its greater flexibility and ability to absorb geometric distortions, yet this gap may narrow with richer component BSDFs \cite{kautz1999interactive}.

DART can be extended in multiple ways.
Examples include adding diffraction modeling to ART \cite{siltanen2008diffraction}. We can also combine other geometric acoustics models, e.g., the image-source method, to mitigate blurred specular components caused by the spatial discretization.
Incorporating learnable delay lines \cite{mezza2024data} could help compensate for inaccurate propagation delays caused by geometric distortion, beyond what the unconstrained material matrices can absorb.
Furthermore, while we introduced various techniques to improve efficiency, they do not fundamentally change the quadratic complexity in the number of radiances, which currently limits DART to geometries with up to around one thousand patches. Reducing this cost would allow handling finer or more complex geometries.

This work focused on echogram prediction, and obtaining RIRs with DART requires an additional auralization step.
We report a preliminary investigation with a simple auralization pipeline in the supplementary material, which shows promising results using the current full-band echogram predictions.
Extending to frequency-dependent modeling is a natural next step that could further improve the auralization quality.

Although DART is a pure signal processing model, it can be combined with neural networks.
For instance, a neural network could predict material parameters from geometry and sparse measurements, enabling cross-scene generalization without per-scene optimization \cite{majumder2022few, liu2025hearing}.
Alternatively, the radiances could be parameterized by a neural network trained to satisfy the ARE \cite{hadadan2022differentiable, hadadan2023inverse}, or a neural network could model the residual of DART's predictions \cite{jin2025avdar, liang2025p}.
Ultimately, the challenge is to develop a model that leverages physical priors when data is scarce, yet scales flexibly to rich, data-driven scenarios.

\appendices

\section{Delay Network Interpretation of ART}\label{app:delay-network}

ART is described with the discrete ARE \eqref{eq:discrete-are}, along with the corresponding discrete injection \eqref{eq:injection} and detection \eqref{eq:detection-main}. It can be viewed as a specific instance of a delay network. In a general form, a delay network can be written as
\begin{subequations}
\begin{align}
    \mathbf{p}[n] &= \mathbf{A}[n] \circledast \mathbf{p}[n] + \mathbf{b}[n] \circledast u[n], \label{eq:delay-net-1} \\
    y[n] &= \mathbf{c}^\top[n] \circledast \mathbf{p}[n] + \mathrm{d}[n] \circledast u[n], \label{eq:delay-net-2}
\end{align}
\end{subequations}
where $u[n]$ and $y[n]$ are scalar input and output signals, and $\mathbf{p}[n]$ is an $N \times 1$ vector of internal states.
$\mathbf{A}[n]$, $\mathbf{b}[n]$, $\mathbf{c}[n]$, and $\mathrm{d}[n]$ correspond to an $N \times N$ feedback mixing filter matrix, $N \times 1$ input filters, $N \times 1$ output filters, and a $1\times 1$ bypass filter. The term ``delay network'' refers to the fact that parallel delay lines are embedded in the feedback recursion $\mathbf{A}[n]$.

ART and DART can be written in this form by identifying the filter components as
\begin{subequations}
\begin{align}
\mathbf{A}[n] &= \hat{\bm{R}}[n] \approx \hat{\bm{M}} \hat{\bm{V}} \diag \hat{\bm{D}}[n], \\
\mathbf{b}[n] &= \hat{\bm{L}}^{(0)}[n] \approx \hat{\bm{M}} (\diag \hat{\bm{G}})^{-1} \hat{\bm{P}}^{\mathrm{(in)}}[n],  \\
\mathbf{c}[n] &= \hat{\bm{W}}[n], \\
\mathrm{d}[n] &= \hat{E}_D[n],
\end{align}
\end{subequations}
where the two approximations are those introduced for DART, e.g., the kernel decomposition (\ref{eq:kernel-decomposition} and \ref{eq:scattering-decomposition}) and injection approximation \eqref{eq:injection-wo-monte-carlo}.
When the input is a unit impulse emission, $u[n] = \delta[n]$, \eqref{eq:delay-net-1} is exactly the same as the discrete ARE \eqref{eq:discrete-are}, and the internal state corresponds to the discrete radiance: $\mathbf{p}[n] = \hat{\bm{L}}[n]$. Applying the output (detection) filter then yields the detected power,
$\mathbf{c}^\top[n] \circledast \mathbf{p}[n] = \hat{E}_L[n]$, same as the discretized detection \eqref{eq:detection-main}. Finally, \eqref{eq:delay-net-2} computes the echogram $y[n] = \hat{E}[n]$.
The number of parallel delay lines is given by $N = N_\mathrm{rad}$ or $N = \bar{N}_\mathrm{rad}$ when exploiting the radiance sparsity, as discussed in Section~\ref{subsec:radiance-computation}.

\section{Derivations}
\subsection{Visibility-Material Decomposition \eqref{eq:scattering-decomposition}} \label{app:kernel-decomposition}
First, we rewrite the scalar-valued matrix \eqref{eq:scattering-kernel} by reparameterizing the source patch integral into a solid angle integral.
\begin{equation}\label{kernel-decomp-proof}
\begin{split}
&\hat{S}_{h j, i k}
=
\iint_{\calS_{i k}} \iint_{\A_i} \iint_{\calS_{h j}} \iint_{\S}  \\
&\,
V_h(x', \Phi)   \rho_i (\Phi, \vout) |\Phi\cdot\nu_i| \delta(\Phi +  \Theta) d\Phi d\Theta \frac{d\A (x')}{|\A_i|} \frac{d\Omega}{|\calS_{i k}|}.
\end{split}
\end{equation}
Note that we also replaced the visibility term $V(x, x')$ with a patch-dependent one $V_h(x', \Phi)$, which denotes ``a ray emitted from $x'$ in direction $\Phi$ hits patch $\A_h$.''
Next, by collapsing the integration over outgoing directions $\Theta \in \S_{hj}$  using the Dirac delta and decomposing the solid angle integral over the sphere into multiple incident directions, i.e., $\S = \cup_l \S_{il}$, we can write
\begin{equation}\label{eq:decomposition-step-2}
\begin{split}
&\hat{S}_{h j, i k} =
\sum_{l=1}^\Ndir \iint_{\calS_{i k}} \iint_{\S_{il}} \\ &\quad\;\Bigg[\iint_{\A_i}
V_{hj}(x', \Phi) \frac{d\A (x')}{|\A_i|} \Bigg]  \rho_i (\Phi, \vout)  |\Phi\cdot\nu_i|  d\Phi \frac{d\Omega}{|\calS_{i k}|}.
\end{split}
\end{equation}
Now, assume that the average visibility (the square-bracketed term) for the incident direction $\Phi$ is approximately constant within each discretized bin $\calS_{i l}$:
\begin{equation}
    \iint_{\A_i} V_{hj}(x', \Phi) \frac{d\A (x')}{|\A_i|} \approx \hat{V}_{hj,il}, \quad \Phi \in \S_{il},
\end{equation}
Then, we can replace the visibility integral in \eqref{eq:decomposition-step-2} with the mean visibility matrix $\smash{\hat{V}_{hj,il}}$ and take it out from the integrals. Under this approximation, the remaining integral reduces to the definition of the material matrix $\smash{\hat{M}_{il,ik}}$, thereby yielding the visibility-material decomposition equation \eqref{eq:scattering-decomposition}.

\subsection{Material Matrix-Based Injection \eqref{eq:injection-wo-monte-carlo}} \label{app:injection}
We can rewrite the discrete initial radiance as
\begin{equation}\label{eq:injection-derivation}
    \begin{split}
    &\hat{L}^{(0)}_{i k}[n]
     =
    \sum_{l=1}^\Ndir \frac{1}{|\A_i|}\iint_{\calS_{i k}} \iint_{\S} \\
    &\; {\Gamma_s(\Theta, o_s)} D(x_s, x, n\Delta t) V_{il}(x_s, \Theta)   \rho_i (-\Theta, \vout) \frac{d\Theta}{4\pi} \frac{d\Omega}{|\calS_{i k}|}
\end{split}
\end{equation}
where we used the solid angle reparameterization and split the input directions into $\S_{il}$ by changing the visibility term to $V_{il}$.

Now, we introduce three approximations.
First, analogous to the simplification step in the reflection kernel \eqref{eq:kernel-decomposition}, we separate the delay term from the integral. Second, we assume that the BSDF $\rho_i$ is approximately constant for incident direction bin $\S_{il}$ and outgoing direction bin $\S_{ik}$. Then, we can separate the material term from the integral.
\begin{equation}
    \hat{L}_{ik}^{(0)}[n]
    \approx
    \sum_{l=1}^\Ndir \hat{D}_{s, il}[n] \Bigg[ \iint_{\S} {\Gamma_s(\Theta, o_s)} V_{il}(x_s, \Theta) \frac{d\Theta}{4\pi} \Bigg] \tilde{M}_{il,ik}, \label{eq:injection-delay-approx}
\end{equation}
where
\begin{equation}
    \tilde{M}_{il,ik} = \frac{1}{|\A_i|} \iint_{\calS_{i k}}  \iint_{\S_{il}}  \rho_i (\Phi, \vout) \frac{d\Phi}{|\S_{il}|} \frac{d\Omega}{|\calS_{i k}|}.   \label{eq:bsdf-integral}
\end{equation}
Finally, we can insert the factor $|\Phi\cdot \nu_i|/|\Phi\cdot \nu_i| = 1$ into the integrand of the material equation \eqref{eq:bsdf-integral} and split its denominator out of the integral using the following approximation:
\begin{equation}
    |\Phi \cdot \nu_i| \approx \iint_{\S_{il}} |\bar{\Phi} \cdot \nu_i | \frac{d \bar{\Phi}}{|\S_{il} |}, \quad {\Phi} \in \S_{il}.
\end{equation}
We then obtain
\begin{equation}
    \tilde{M}_{il,ik} \approx \hat{M}_{il,ik} / \hat{G}_{il} .
\end{equation}
Combining it with \eqref{eq:injection-delay-approx} gives the discretized injection equation with the material matrix \eqref{eq:injection-wo-monte-carlo}.

\section{Aliasing Suppression}\label{app:aliasing}

Frequency-sampling the $z$-domain ARE \eqref{eq:discrete-are-frequency} at $z = z_{T,f}$, i.e., $T$ uniformly spaced points on the unit circle, and applying the IFFT yields length-$T$ signals that suffer from time-aliasing~\cite{smith2007mathematics}. For the exact radiance $\smash{\hat{\bm{L}}[n]}$, its frequency-sampled, hence aliased, signal is given as
\begin{equation}\label{eq:aliasing}
  \hat{\bm{L}}_T[n]
  = \sum_{m=0}^{\infty} \hat{\bm{L}}[n + mT],
  \quad 0 \le n < T.
\end{equation}
The total aliasing error is
\begin{equation}\label{eq:aliasing_error}
  \mathcal{E}_T
  = \bigl\| \hat{\bm{L}}_T[n]
    - \hat{\bm{L}}[n] \bigr\|
  = \left\| \sum\nolimits_{m=1}^{\infty}
    \hat{\bm{L}}[n + mT] \right\|\!,
\end{equation}
where $\|\cdot\|$ denotes any $p$-norm ($p \ge 1$) over the first $T$ samples.
Because the radiance signals are nonnegative, aliasing always causes overestimation, with pronounced errors at the beginning of the signal where the true energy is zero.

To suppress the aliasing, we frequency-sample on a circle of radius $\gamma^{-1/T} > 1$. Since $z_{T,\gamma,f} = \gamma^{-1/T} z_{T,f}$,
frequency-sampling at $\smash{z_{T,\gamma,f}}$ is equivalent to sampling the original signal damped with an exponentially decreasing window $\gamma^{n/T}$ in the time domain.
Therefore, applying the IFFT and compensating the damping with an exponentially increasing window $\gamma^{-n/T}$ yields the aliasing-suppressed signal
\begin{subequations}\label{eq:damped_signal}
\begin{align}
\hat{\bm{L}}_{T,\gamma}[n] & = \gamma^{-n/T}\sum_{m=0}^{\infty} \gamma^{n/T + m}\,
\hat{\bm{L}}[n + mT]\\
& = \sum_{m=0}^{\infty} \gamma^{m}\,
\hat{\bm{L}}[n + mT].
\end{align}
\end{subequations}
The time-aliasing error therefore satisfies
\begin{equation}\label{eq:damped_error}
  \mathcal{E}_{T,\gamma}
  = \left\| \sum\nolimits_{m=1}^{\infty}
     \gamma^{m}\,
     \hat{\bm{L}}[n + mT] \right\|
  < \gamma\, \mathcal{E}_T,
\end{equation}
where the inequality uses nonnegativity of the radiance and
each aliased tail ($m \ge 1$) being scaled by $\gamma^m < 1$.

We assumed the exact radiance for the above derivation, but a similar bound holds for the truncated Neumann series.
Furthermore, since high-order terms predominantly contribute to the tail $n \geq T$, the error of the truncated series $\smash{\mathcal{E}_{T,N_\mathrm{order}}}$ or its antialiased version $\smash{\mathcal{E}_{T,N_\mathrm{order},\gamma}}$ is no larger than the original counterpart ($\mathcal{E}_{T}$ and $\mathcal{E}_{T,\gamma}$, respectively).
In other words, increasing the truncation order beyond what is needed to capture the dominant energy within the echogram duration $T$ yields no benefit, as the additional terms only add to the aliased tail.

\bibliography{refs}
\bibliographystyle{IEEEtran}

\newpage
\end{document}